\documentclass[11pt]{article}
\usepackage{jheppub_modified_green}

\usepackage{tocloft}
\usepackage{epsfig}
\usepackage{amsfonts}
\usepackage{latexsym}
\usepackage{amsmath,amssymb}
\usepackage{graphicx}

\newcommand{\nc}{\newcommand}
\nc{\rnc}{\renewcommand}
\rnc{\d}{\mathrm{d}}
\nc{\D}{\partial}
\nc{\K}{\kappa}
\nc{\bK}{\bar{\K}}
\nc{\bN}{\bar{N}}
\nc{\bq}{\bar{q}}
\nc{\bp}{\bar{p}}
\nc{\vbq}{\vec{\bar{q}}}
\nc{\g}{\gamma}
\nc{\lrarrow}{\leftrightarrow}
\nc{\rg}{\sqrt{g}}
\rnc{\[}{\begin{equation}}
\rnc{\]}{\end{equation}}
\nc{\bea}{\begin{eqnarray}}
\nc{\eea}{\end{eqnarray}}
\nc{\nn}{\nonumber}
\rnc{\(}{\left(}
\rnc{\)}{\right)}
\nc{\q}{\vec{q}}
\nc{\x}{\vec{x}}
\nc{\ep}{\epsilon}
\nc{\tto}{\rightarrow}
\rnc{\inf}{\infty}
\rnc{\Re}{\mathrm{Re}}
\rnc{\Im}{\mathrm{Im}}
\nc{\z}{\zeta}
\nc{\mA}{\mathcal{A}}
\nc{\mB}{\mathcal{B}}
\nc{\mC}{\mathcal{C}}
\nc{\mD}{\mathcal{D}}
\nc{\mE}{\mathcal{E}}
\nc{\mF}{\mathcal{F}}
\rnc{\H}{\mathcal{H}}
\rnc{\L}{\mathcal{L}}
\nc{\<}{\langle}
\rnc{\>}{\rangle}
\nc{\fnl}{f_{NL}}
\nc{\fnleq}{f_{NL}^{equil.}}
\nc{\fnlloc}{f_{NL}^{local}}
\nc{\vphi}{\varphi}
\nc{\Lie}{\pounds}
\nc{\half}{\frac{1}{2}}
\nc{\bOmega}{\bar{\Omega}}
\nc{\bLambda}{\bar{\Lambda}}
\nc{\dN}{\delta N}
\nc{\gYM}{g_{\mathrm{YM}}}
\nc{\geff}{g_{\mathrm{eff}}}
\nc{\bg}{\gamma} 
\nc{\Oi}{\Omega_{[2]}}
\nc{\Oii}{\Omega_{[3]}}
\nc{\Ei}{E_{[2]}}
\nc{\Eii}{E_{[3]}}
\nc{\bOi}{\bar{\Omega}_{[2]}}
\nc{\bOii}{\bar{\Omega}_{[3]}}
\nc{\bEi}{\bar{E}_{[2]}}
\nc{\bEii}{\bar{E}_{[3]}}
\rnc{\a}{\bar{a}}
\rnc{\b}{\bar{b}}
\rnc{\c}{\bar{c}}
\rnc{\O}{\mathcal{O}}
\nc{\blambda}{\bar{\lambda}}
\nc{\oa}{\stackrel{\leftrightarrow}}

\newcommand{\p}{\partial}
\nc{\wT}{\widetilde{T}}
\rnc{\ll}{\<\!\<}
\nc{\rr}{\>\!\>}
\nc{\Y}{\Upsilon}
\nc{\G}{\mathcal{G}}

\title{Soft limits in holographic cosmology} 
\author{Paul McFadden}
\affiliation{Theoretical Physics Group, Blackett Laboratory, Imperial College London, SW7 2AZ, UK.\\[0.1ex]
Perimeter Institute for Theoretical Physics, Waterloo Ontario, Canada N2L 2Y5.}
\emailAdd{p.mcfadden@imperial.ac.uk}
\abstract{
We study the soft limits of cosmological correlators
from a holographic perspective, 
showing how the inflationary consistency relations arise from the diffeomorphism invariance of the dual quantum field theory.
Starting from the corresponding Ward identity, by taking moments
we derive the leading and subleading behaviour of the stress tensor 3-point function in the limit as one momentum vanishes. 
These results are non-perturbative and valid in quantum field theories of a very general nature.
Exploiting the known mapping of correlators 
in the dual quantum field theory
to those of the cosmology, we then obtain the leading and subleading soft behaviour of all cosmological 3-point correlators of curvature perturbations and gravitons.
Our results thus provide a holographic derivation of all leading and subleading consistency relations for cosmological 3-point functions, 
and our method is easily generalised. 
We verify our results explicitly for slow-roll inflation and 
for strongly coupled holographic cosmologies 
with a perturbative dual description.

}
\arxivnumber{}

\begin{document}

\maketitle
%\thispagestyle{empty}
%\newpage
%\addtocounter{page}{-1}

\section{Introduction}

The correlators of primordial perturbations 
encode crucial clues to the dynamics of the early universe.  
Of particular importance are the soft limits of these correlators, in which one or more of the momenta vanish.
In such limits, it is possible to derive exact non-perturbative statements that are largely model independent and rely on only a few broad dynamical assumptions. 
The most celebrated of these are the inflationary consistency conditions \cite{Maldacena:2002vr, Creminelli:2004yq, Cheung:2007sv,
Creminelli:2011rh, Creminelli:2012ed, Creminelli:2013cga, Hinterbichler:2013dpa} which relate $n$-point correlators in the limit where one momentum vanishes to $(n{-}1)$-point functions.
Valid in any single-field model for which the background is an attractor \cite{Creminelli:2004yq}, any observed  violation, 
whether in the cosmic microwave background \cite{Creminelli:2011sq, Bartolo:2011wb, Pajer:2013ana, Ade:2013ydc} or in large-scale structure \cite{Giddings:2011zd, Dai:2013kra, Kehagias:2013yd, Peloso:2013zw, Creminelli:2013mca, Horn:2014rta, Dimastrogiovanni:2014ina, Alvarez:2014vva},
would be a signature of more exotic 
dynamics (e.g., multiple fields and/or
the growth of curvature perturbations outside the horizon \cite{Byrnes:2010em, Suyama:2013nva}, non-attractor behaviour \cite{Namjoo:2012aa, Martin:2012pe, Chen:2013aj}, or departures from the Bunch-Davies vacuum \cite{Chen:2006nt, Holman:2007na, Ashoorioon:2010xg, Agullo:2010ws, Ganc:2011dy, Ashoorioon:2013eia, Flauger:2013hra, Berezhiani:2014eia, Collins:2014fwa}). 
Given their obvious importance, the consistency relations have been studied from a variety of standpoints 
including background-wave arguments \cite{Creminelli:2004yq, Cheung:2007sv, Creminelli:2011rh, Creminelli:2012ed, Creminelli:2013cga}, the wavefunction of the universe \cite{Pimentel:2013gza},
and the symmetries of adiabatic perturbation modes 
\cite{Hinterbichler:2012nm, Hinterbichler:2013dpa, Berezhiani:2013ewa, Goldberger:2013rsa, Berezhiani:2014tda, Armendariz-Picon:2014xda}.

Our aim in this paper is to understand the consistency relations 
from a holographic perspective.
In holographic cosmology, inflationary 
correlators are determined by the 
stress tensor correlators of a
dual quantum field theory (QFT), which is both three dimensional and non-gravitational.
The cosmological consistency relations should therefore be equivalent to  
soft theorems relating $n$-point stress tensor correlators in the limit as one momentum vanishes to $(n{-}1)$-point correlators.
In effect, holography should reduce the analysis of cosmological soft limits to a straightforward problem in ordinary QFT.

Despite originating in the same paper as the consistency relations themselves \cite{Maldacena:2002vr}, this promising 
perspective has yet to be systematically developed.
Partial progress was made in \cite{Schalm:2012pi}, which recovered the leading order consistency relations for the scalar bispectrum using the Callan-Symanzik equation, and \cite{Bzowski:2012ih}, which used conformal perturbation theory to recover the leading consistency relations for both the scalar bispectrum and the 3-point function of a soft curvature perturbation and two gravitons. 
Both these works required however specific assumptions about the nature of the dual QFT (either pertaining to the form the $\beta$-function or proximity to a fixed point) that are considerably more restrictive than 
those required to derive the bulk cosmological consistency relations.
Moreover, it is not clear how either approach can be extended 
to obtain consistency relations for soft gravitons, or to understand soft behaviour at subleading orders.

In this paper we propose instead a fresh approach that eliminates 
these difficulties. This approach is non-perturbative and 
applies under conditions equivalent to those assumed in the cosmology.
Besides enabling a holographic derivation of the consistency relations for soft curvature perturbations, we can handle soft gravitons and extract the complete subleading soft behaviour.  In fact, 
as we will show in a companion paper \cite{toappear}, 
it is even possible to  
recover the entire infinite hierarchy of consistency relations discovered in \cite{Hinterbichler:2013dpa}, although in this paper we will confine ourselves to an analysis 
of the leading and subleading soft behaviour. 
On top of these advantages, the method is both 
simple and systematic. 

Our starting point is the diffeomorphism Ward identities in the dual QFT.  These may be obtained by functionally differentiating with respect to the metric the generating relation,
\[\label{genrel}
0 = \nabla^i \<T_{ij}(\x)\>_s,
\]
before restoring a flat metric.  (Here, the subscript $s$ on the 1-point function indicates the presence of a non-zero source, namely, a non-flat metric.) 
As required, these are exact and non-perturbative statements relating the divergence of $n$-point functions of the stress tensor to contact terms involving $(n{-}1)$-point functions (or lower).  
The detailed form of these Ward identities depends 
on only two assumptions.  Firstly, that 1-point functions in the absence of sources vanish,\footnote{This is true when the dual QFT is in the Euclidean vacuum state, as corresponds to Bunch-Davies initial conditions in cosmology \cite{McFadden:2009fg}, but might be violated more generally.} and secondly, that only a single bulk scalar is present.  This latter condition 
ensures that the source for the dual scalar operator is spatially uniform in the background, eliminating 
an additional contribution to \eqref{genrel} from the scalar 1-point function in the presence of sources.\footnote{For multiple bulk scalars such contributions generically arise, however, as we discuss in section \ref{discussion}. 
These contributions lead to Ward identities with both pure stress tensor and mixed stress tensor/scalar $n$-point functions, blocking our derivation of the consistency relations as required.}

To derive soft theorems for stress tensor correlators, we simply take moments of these diffeomorphism Ward identities.  Multiplying both sides by $x_a x_b\ldots$ and integrating over all $\x$, on the right-hand side we pick up a finite number of $(n{-}1)$-point contributions from integrating over the contact terms, while the left-hand side can be handled by parts yielding (modulo a boundary term) the required moment integrals, e.g.,
\[
\int \d^3 \x\,  x_a x_b\, \p_i \<T_{ij}(\x)\ldots\>
= -2\int \d^3\x \,\delta_{i(a}x_{b)}\<T_{ij}(\x)\ldots\>.
\]
In momentum space, these become soft theorems for the zero-momentum limit of momentum-derivatives of stress tensor correlators.  (One could alternatively work in momentum space throughout by differentiating the Fourier transform of the Ward identity with respect to the soft momentum.)
One can then straightforwardly reconstruct the Taylor expansion for the leading and subleading soft behaviour of the stress tensor $n$-point function.

To derive the cosmological consistency relations from these soft theorems, we simply use the holographic formulae linking stress tensor correlators of the dual QFT to cosmological correlators.  In this paper we focus on 3-point correlators, since the required holographic formulae (both for scalars and tensors) have been completely worked out in this case \cite{McFadden:2010vh, McFadden:2011kk}.\footnote{The attractor property of the background is implicit in the derivation of these holographic formulae, which assume the dual QFT either flows to a fixed point or has generalised conformal symmetry in the UV.}
The soft limit we study is therefore the familiar squeezed limit of the 3-point function. 
Specifically, we will show how to recover the leading and subleading soft behaviour (i.e., to $O(q_1^2)$) of all cosmological 3-point correlators of both scalars and tensors.  Included in these results are the consistency relations of \cite{Maldacena:2002vr, Creminelli:2011rh, Creminelli:2012ed} plus the lower-order relations of \cite{Hinterbichler:2013dpa}.
Our method is readily extendible to $n$-point functions, however, upon determination of the appropriate holographic formulae. 
This would allow a holographic analysis of internal \cite{Senatore:2012wy, Assassi:2012zq} and multiple soft limits \cite{Mirbabayi:2014zpa, Joyce:2014aqa}.

Since the soft theorems we derive for stress tensor correlators are non-perturbative, they hold equally well when the dual QFT is strongly or weakly coupled.  The latter case is especially interesting since it corresponds to an early universe emerging from a strongly coupled holographic phase \cite{McFadden:2009fg,  McFadden:2010na, McFadden:2010jw, McFadden:2010vh, Easther:2011wh, Bzowski:2011ab, Kawai:2014vxa}.   As the string scale is comparable to the Hubble scale in this case (though both are far below the Planck scale), a geometric description in terms of low-energy fields such as the metric is not straightforwardly applicable\footnote{Indeed, the possibility this entails of avoiding the big bang singularity is a key motivation for these models. The absence of a big bang singularity 
corresponds to IR finiteness of the dual QFT, as discussed in \cite{McFadden:2010na, Easther:2011wh}.} meaning that the  standard methods for deriving cosmological consistency relations cannot be applied.  Instead one can follow the holographic approach we develop here, working directly in the dual QFT.
The validity of the consistency relations even under these extreme circumstances goes a long way towards explaining the remarkable fact that, despite the very different underlying physical picture,  
the predicted correlators of this holographic model are still very close to those of ordinary slow-roll inflation \cite{Bzowski:2011ab}.

The layout of this paper is as follows.  In section \ref{sec_soft} we define the soft limit and introduce the diffeomorphism Ward identity.   Next, we derive the soft theorems associated with the first and second moments of this Ward identity.  Using these results, we then reconstruct the leading and subleading soft behaviour of the stress tensor 3-point function in a convenient helicity basis.
A useful check of these results follows from previous calculations for free fields.  (Note that we work in three dimensions throughout, though our results are easily generalised.)
In section \ref{sec_hc}, we introduce the holographic formulae linking QFT correlators to cosmological correlators.  We then recover all leading and subleading consistency relations for cosmological 3-point correlators by inserting our results for the soft limits of the stress tensor 3-point function.
In section \ref{sec_tests}, we test these consistency relations for standard slow-roll inflation and also strongly coupled holographic cosmologies based on a perturbative dual QFT description.  Possible generalisations of our results are noted in the discussion section, as well as some preliminary considerations of how violations of the consistency relations show up in the dual QFT description.  Three appendices contain additional technical information.  Appendix \ref{higher_app} discusses 
the convergence of the boundary term when computing moments of the Ward identity.  Appendix \ref{Fourier_app} provides additional detail relating to Fourier transforms as well as a direct momentum-space derivation of the soft theorems.  Finally, appendix \ref{helicity_basis_app} lists our conventions and a few useful properties of polarisation tensors.

\section{Soft theorems for stress tensor correlators}\label{sec_soft}

\subsection{Defining the soft limit}

Let us begin by defining carefully what we mean by the soft (or squeezed) limit of a 3-point function.  
In this limit the magnitude of one of the momenta, say $\q_1$, is taken to zero while preserving its direction.  
To fully specify the limit, we additionally need to prescribe what happens to the two remaining momenta $\q_2$ and $\q_3$.

One possibility would be to select one of these as our preferred momentum and hold it fixed while sending $q_1\tto 0$, with the remaining momentum following from overall momentum conservation.  
To treat $\q_2$ and $\q_3$ on a more equal footing, however, we prefer instead to hold fixed the antisymmetric linear combination,\footnote{The symmetric linear combination is fixed by momentum conservation to be $-\frac{1}{2}\q_1$.}
\[
\q = \frac{1}{2}(\q_2-\q_3),
\]
in line with some of the cosmological literature \cite{Creminelli:2004yq, Creminelli:2011rh, Creminelli:2012ed}. We emphasize that this is simply a choice, however, and that our method works equally well for other choices.
 
\begin{figure}
\center
\includegraphics[width=5.8cm]{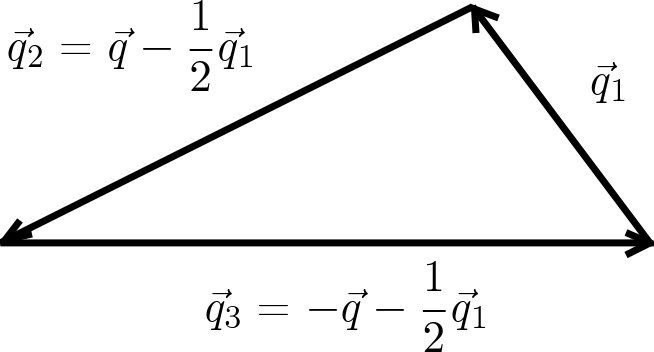}
\vspace{0.1cm} %5.8cm
\caption{\label{sq_fig}
Due to momentum conservation the momenta in a 3-point function form a triangle.  In the squeezed limit we send the magnitude of $\q_1$ to zero while holding its direction fixed.  The remaining vectors $\q_2$ and $\q_3$ are then given in terms of $\q_1$ and the fixed vector $\q$ as shown.
}
\end{figure}

Our soft limit thus corresponds to setting 
\[\label{q2q3}
\q_2 = \q -\frac{1}{2}\q_1, \qquad \q_3 = -\q -\frac{1}{2}\q_1,
\]
then sending $q_1\tto 0$ while holding $\q$ and $\q_1/q_1$ fixed, as illustrated in Fig.~\ref{sq_fig}.
In this limit, scalar quantities can be expressed as functions of the magnitudes $q$ and $q_1$, along with the fixed angle $\vphi$ defined by
\[\label{vphi_def}
\cos\vphi = \frac{\q_1\cdot \q}{q_1 \,q}.
\]
Similarly, when we convert to a helicity basis in section \ref{hel_conv}, all our results will be expressible in terms of $q$, $q_1$ and $\vphi$.

\subsection{The diffeomorphism Ward identity}

Returning now to position space, as noted above, the diffeomorphism Ward identity for stress tensor correlators may be obtained by functionally differentiating the basic identity $\nabla^i\<T_{ij}(\x)\>_s=0$ 
with respect to the metric.  After this, we restore the metric to be flat and we can write all indices as lowered.

Functionally differentiating once, we find the well-known 2-point Ward identity
\[\label{2ptWard}
\p_i\<T_{ij}(\x_1)T_{kl}(\x_2)\>=0.
\] 
Functionally differentiating twice yields the 3-point Ward identity,
\begin{align}
\label{WardId}
& \p_i \Big[ \<T_{ij}(\x_1)T_{kl}(\x_2)T_{mn}(\x_3)\>\nn\\&\qquad
-2\<T_{ij}(\x_1)\Y_{klmn}(\x_2,\x_3)\> 
-2\<\Upsilon_{ijkl}(\x_1,\x_2)T_{mn}(\x_3)\>-2\<\Upsilon_{ijmn}(\x_1,\x_3)T_{kl}(\x_2)\>\Big]\nn\\[1ex]
& = 2\p_{(k}\big(\<T_{l)j}(\x_1)T_{mn}(\x_3)\>\delta(\x_2-\x_1)\big)
+2\p_{(m}\big(\<T_{n)j}(\x_1)T_{kl}(\x_2)\>\delta(\x_3-\x_1)\big)\nn\\[1ex]&\quad
-\delta_{kl}\<T_{ij}(\x_1)T_{mn}(\x_3)\>\p_i\delta(\x_2-\x_1)
-\delta_{mn}\<T_{ij}(\x_1)T_{kl}(\x_2)\>\p_i\delta(\x_3-\x_1)\nn\\[1ex]&\quad
+\<T_{kl}(\x_1)T_{mn}(\x_3)\>\p_j\delta(\x_2-\x_1)
+\<T_{mn}(\x_1)T_{kl}(\x_2)\>\p_j\delta(\x_3-\x_1).
\end{align}
In both these relations all partial derivatives are taken with respect to $\x_1$, and we drop ultralocal contact terms that only contribute when all three insertion points coincide. (Such terms depend on the choice of renormalisation scheme and can be removed by the addition of local counterterms.)
  The tensor $\Upsilon_{ijkl}$ encodes the residual metric-dependence hidden within the stress tensor $T_{ij}$ itself, and is defined by
\[
\Upsilon_{ijkl}(\x_1,\x_2) = \frac{\delta T_{ij}(\x_1)}{\delta g^{kl}(\x_2)}\Big|_0
\]
where the zero subscript indicates the sources have been set to zero (i.e., the metric restored to flatness).
In \eqref{WardId} we have defined the 3-point function to mean the insertion of three copies of the stress tensor $T_{ij}$.  An alternative definition (used, for example, in \cite{Osborn:1993cr}) would be to functionally differentiate the generating functional three times.  This latter definition effectively subsumes the terms involving $\Y_{ijkl}$ into the stress tensor 3-point function itself, eliminating them from the Ward identity.  Which definition is used is purely a matter of convention, however, and it is straightforward to convert between the two by the addition of the appropriate semilocal terms (i.e., terms which contribute when only two of the three insertions points are coincident).

Notice also that, despite appearances, the first term on the second line of \eqref{WardId} is actually symmetric under interchange of $\x_2$ and $\x_3$, i.e.,
\[
\p_i \<T_{ij}(\x_1)\Y_{klmn}(\x_2,\x_3)\> = \p_i\<T_{ij}(\x_1)\Y_{mnkl}(\x_3,\x_2)\>.
\]
To see this, going back to the definition of the stress tensor in terms of the action, one finds that \cite{McFadden:2011kk} 
\[\label{Yswap}
\Y_{klmn}(\x_2,\x_3) = \Y_{mnkl}(\x_3,\x_2)+\frac{1}{2}\big(T_{kl}(\x_2)\delta_{mn}-T_{mn}(\x_3)\delta_{kl}\big)\delta(\x_2-\x_3).
\]
The 2-point Ward identity \eqref{2ptWard} then eliminates the non-symmetric piece when inserted into the correlator.
In fact, for free fields, the entire term $\p_i\<T_{ij}(\x_1)\Y_{klmn}(\x_2,\x_3)\>$ vanishes since $\Y_{ijkl}$ can always be rewritten purely in terms of the stress tensor \cite{Bzowski:2011ab}. (For this reason this term is omitted from the Ward identity quoted in Appendix D.3 of \cite{Bzowski:2011ab}.)  As we consider completely general QFTs here, however, we retain this term explicitly.

\subsection{Moments of the diffeomorphism Ward identity}

\subsubsection{First moment}

To derive the leading-order behaviour in the squeezed limit, we take the first moment of the 3-point Ward identity \eqref{WardId}.
To do this, we multiply  by $x_{1a}$ then integrate over all $\x_1$.  
We obtain a contribution from each of the contact terms on the right-hand side,
while the left-hand side is handled by parts to eliminate the partial derivative.\footnote{The boundary term vanishes, though for higher moments this is non-trivial as discussed in appendix \ref{higher_app}.}
In this fashion, we obtain
\begin{align}\label{low1}
& \int\d^3\x_1\,\Big[\< T_{ij}(\x_1)T_{kl}(\x_2)T_{mn}(\x_3)\>
\nn\\&\quad
-2\<T_{ij}(\x_1)\Y_{klmn}(\x_2,\x_3)\>
-2\<\Y_{ijkl}(\x_1,\x_2)T_{mn}(\x_3)\>-2\<\Y_{ijmn}(\x_1,\x_3)T_{kl}(\x_2)\>\Big]\nn\\[1ex]&
=
2\delta_{i(k}\< T_{l)j}(\x_2)T_{mn}(\x_3)\>
+2\delta_{i(m}\< T_{n)j}(\x_3)T_{kl}(\x_2)\> 
+2\delta_{ij} \< T_{kl}(\x_2)T_{mn}(\x_3)\> \nn\\[1ex]&\quad
-\delta_{kl}\< T_{ij}(\x_2)T_{mn}(\x_3)\>
-\delta_{mn}\< T_{ij}(\x_3)T_{kl}(\x_2)\> 
+\Big(x_{2i}\frac{\p}{\p x_{2j}}+x_{3i}\frac{\p}{\p x_{3j}}\Big)\<T_{kl}(\x_2)T_{mn}(\x_3)\>.
\end{align}
In momentum space, the integral over $\x_1$ on the left-hand side corresponds to an insertion at zero momentum, with momentum conservation then forcing the remaining momenta to be equal and opposite ($\pm\q$ from \eqref{q2q3}).  Thus, after Fourier transforming,
we obtain the result\footnote{See appendix \ref{Fourier_details} for details, and appendix
 \ref{momsp} for a parallel discussion starting from the momentum-space Ward identity.}
\begin{align}\label{soft1}
& \lim_{q_1\tto 0}\Big[\ll T_{ij}(\q_1)T_{kl}(\q_2)T_{mn}(\q_3)\rr
\nn\\&\quad
-2\ll T_{ij}(\q_1)\Y_{klmn}(\q_2,\q_3)\rr
-2\<\!\<\Upsilon_{ijkl}(\q_1,\q_2)T_{mn}(\q_3)\>\!\> -2\<\!\<\Upsilon_{ijmn}(\q_1,\q_3)T_{kl}(\q_2)\>\!\> \Big]
\nn\\[1ex]&
=2\delta_{i(k}\<\!\<T_{l)j}(\q)T_{mn}(-\q)\>\!\>
+2\delta_{i(m}\<\!\<T_{n)j}(\q)T_{kl}(-\q)\>\!\> 
+\delta_{ij} \ll T_{kl}(\q)T_{mn}(-\q)\rr \nn\\&\quad
-\delta_{kl}\ll T_{ij}(\q)T_{mn}(-\q)\rr
-\delta_{mn}\ll T_{ij}(\q)T_{kl}(-\q)\rr 
-q_j\frac{\p}{\p q_i}\ll T_{kl}(\q)T_{mn}(-\q)\rr .
\end{align}
Our double bracket notation $\ll \ldots \rr$
for correlators  here simply indicates the removal of the overall momentum-conserving delta function, i.e.,
\begin{align}\label{chevron}
\<T_{ij}(\q_1)T_{kl}(\q_2)\> &= \ll T_{ij}(\q_1)T_{kl}(-\q_1)\rr (2\pi)^3 \delta(\q_1+\q_2), \\[1ex]
\<T_{ij}(\q_1)T_{kl}(\q_2)T_{mn}(\q_3)\> &= \ll T_{ij}(\q_1)T_{kl}(\q_2)T_{mn}(\q_3)\rr (2\pi)^3\delta(\q_1+\q_2+\q_3),
\end{align}
and similarly for other correlators.  

The first moment of the 3-point Ward identity \eqref{WardId} thus yields the soft theorem \eqref{soft1} relating the leading behaviour of the 3-point function in the squeezed limit to the 2-point function and its derivative.
Applying similar arguments to the 2-point Ward identity \eqref{2ptWard} we also find
\[\label{2pt1st}
\lim_{q_1\tto 0}\ll T_{ij}(\q_1)T_{kl}(-\q_1)\rr = 0.
\]

\subsubsection{Second moment}

To obtain the subleading behaviour of the 3-point function in the squeezed limit, we must instead take the second moment of the Ward identity \eqref{WardId}.  To do this, we multiply both sides by $x_{1a} x_{1b}$ and then integrate over $\x_1$ as before, yielding
\begin{align}\label{low2}
&\int\d^3\x_1\,x_{1(a}\Big[\< T_{b)j}(\x_1)T_{kl}(\x_2)T_{mn}(\x_3)\>\nn\\&\quad
-2\<T_{b)j}(\x_1)\Y_{klmn}(\x_2,x_3)\>
-2\<\Y_{b)jkl}(\x_1,\x_2)T_{mn}(\x_3)\>-2\<\Y_{b)jmn}(\x_1,\x_3)T_{kl}(\x_2)\>\Big]\nn\\[2ex]
&=
2x_{2(a}\delta_{b)(k}\<T_{l)j}(\x_2)T_{mn}(\x_3)\>
+2x_{3(a}\delta_{b)(m}\<T_{n)j}(\x_3)T_{kl}(\x_2)\>
-\delta_{kl}x_{2(a}\<T_{b)j}(\x_2)T_{mn}(\x_3)\>
\nn\\[2ex]&\quad
-\delta_{mn}x_{3(a}\<T_{b)j}(\x_3)T_{kl}(\x_2)\>
+x_{2(a}\delta_{b)j}\<T_{kl}(\x_2)T_{mn}(\x_3)\>
+x_{3(a}\delta_{b)j}\<T_{mn}(\x_3)T_{kl}(\x_2)\>
\nn\\[2ex]&\quad
+\Big(x_{2a}x_{2b}\frac{\p}{\p x_{2j}}+x_{3a}x_{3b}\frac{\p}{\p x_{3j}}\Big)\<T_{kl}(\x_2)T_{mn}(\x_3)\>.
\end{align}
Transforming to momentum space,\footnote{Again, see appendix \ref{Fourier_details} for details.}  one then obtains the subleading soft theorem
\begin{align}\label{subsoft}
\lim_{q_1\tto 0} \,& \frac{\p}{\p q_{1(a}} 
\Big[ \ll T_{b)j}(\q_1)T_{kl}(\q_2)T_{mn}(\q_3)\rr \nn\\&\quad
-2\ll T_{b)j}(\q_1)\Y_{klmn}(\q_2,\q_3)\rr
-2\ll \Y_{b)jkl}(\q_1,\q_2)T_{mn}(\q_3)\rr 
-2\ll \Y_{b)jmn}(\q_1,\q_3)T_{kl}(\q_2)\rr \Big] \nn\\&
= \frac{\p}{\p q_{(a}}\Big[\delta_{b)(k}\ll T_{l)j}(\q)T_{mn}(-\q)\rr
-\delta_{b)(m}\ll T_{n)j}(\q)T_{kl}(-\q)\rr \nn\\&\qquad\qquad\qquad
+\frac{1}{2}\ll T_{b)j}(\q)T_{kl}(-\q)\rr\delta_{mn}
-\frac{1}{2}\ll T_{b)j}(\q)T_{mn}(-\q)\rr\delta_{kl}
\Big],
\end{align}
where the limit $q_1\tto 0$ is taken while enforcing \eqref{q2q3}.

The final step is now to remove the symmetrisation over the indices $a$ and $b$ in \eqref{subsoft}.   
%This can be done using the symmetry of $T_{ij}$ and $\Y_{ijkl}$ under the permutation of $i$ and $j$.  
%Given a symmetric tensor such that 
Noting that for a tensor obeying $X_{abj}=X_{ajb}$, it follows that $X_{abj}=X_{(ab)j}+X_{(aj)b}-X_{(jb)a}$, we find
%In this manner, we find
%{\allowdisplaybreaks
\begin{align}\label{soft2}
\lim_{q_1\tto 0} \,& \frac{\p}{\p q_{1a}} 
\Big[ \ll T_{bj}(\q_1)T_{kl}(\q_2)T_{mn}(\q_3)\rr \nn\\&\quad
-2\ll T_{bj}(\q_1)\Y_{klmn}(\q_2,\q_3)\rr
-2\ll \Y_{bjkl}(\q_1,\q_2)T_{mn}(\q_3)\rr 
-2\ll \Y_{bjmn}(\q_1,\q_3)T_{kl}(\q_2)\rr \Big] \nn\\&
= \,\,\, \frac{1}{2}\frac{\p}{\p q_a}\Big[\delta_{mn}\ll T_{bj}(\q)T_{kl}(-\q)\rr 
-\delta_{kl}\ll T_{bj}(\q)T_{mn}(-\q)\rr \Big] \nn\\&\quad 
+ \frac{\p}{\p q_{(a}}\Big[\delta_{b)(k}\ll T_{l)j}(\q)T_{mn}(-\q)\rr 
- \delta_{b)(m}\ll T_{n)j}(\q)T_{kl}(-\q)\rr \Big]\nn\\&\quad
+ \frac{\p}{\p q_{(a}}\Big[\delta_{j)(k}\ll T_{l)b}(\q)T_{mn}(-\q)\rr 
- \delta_{j)(m}\ll T_{n)b}(\q)T_{kl}(-\q)\rr \Big]\nn\\&\quad
+ \frac{\p}{\p q_{(j}}\Big[\delta_{b)(k}\ll T_{l)a}(\q)T_{mn}(-\q)\rr 
- \delta_{b)(m}\ll T_{n)a}(\q)T_{kl}(-\q)\rr \Big].
\end{align}
%}
This soft theorem relates the subleading behaviour of the momentum-space 3-point function in the squeezed limit (i.e., the terms linear in $\q_1$) to single derivatives of the 2-point function.

Using similar methods to compute the second moment of the 2-point Ward identity \eqref{2ptWard}, we find
\[\label{2pt2nd}
\lim_{q_1\tto 0}\frac{\p}{\p q_{1a}}\ll T_{bj}(\q_1)T_{kl}(-\q_1)\rr = 0.
\]
Generally, we can combine the information contained in the leading and subleading soft theorems to reconstruct the Taylor expansion for correlators in the soft limit.  Thus, 
using \eqref{2pt1st} and \eqref{2pt2nd}, for the 2-point function we have
\begin{align}\label{2ptTaylor}
\ll T_{ij}(\q_1)T_{kl}(-\q_1)\rr & = \lim_{q_1\tto 0}\ll T_{ij}(\q_1)T_{kl}(-\q_1)\rr
+ q_{1a} \lim_{q_1\tto 0}\frac{\p}{\p q_{1a}}\ll T_{bj}(\q_1)T_{kl}(-\q_1)\rr +O(q_1^2)\nn\\&
= O(q_1^2),
\end{align}
while for the 3-point function,
\begin{align}\label{Taylor}
& \ll T_{ij}(\q_1)T_{kl}(\q_2)T_{mn}(\q_3)\rr +\ldots 
\nn\\[1ex]&\quad
=
\lim_{q_1\tto 0}\Big[ \ll T_{ij}(\q_1)T_{kl}(\q_2)T_{mn}(\q_3)\rr+\ldots\Big] %\nn\\[1ex]&\qquad\qquad\qquad
+ q_{1a}\lim_{q_1\tto 0}\frac{\p}{\p q_{1a}}\Big[\ll T_{ij}(\q_1)T_{kl}(\q_2)T_{mn}(\q_3)\rr + \ldots\Big]
\nn\\[1ex]&\qquad
+ O(q_1^2),
\end{align}
where \eqref{soft1} and \eqref{soft2} should be used to evaluate the right-hand side.

\subsection{Converting to a helicity basis}\label{hel_conv}

We now convert our results to a helicity basis by contracting with polarisation tensors and taking traces.
Expressed in this form, our results for the squeezed limit of the stress tensor 3-point function are ready to be fed into our holographic formulae for cosmological correlators.
As we will see later, these holographic formulae associate trace components of the stress tensor with cosmological curvature perturbations, while helicity components are associated with gravitons.

We begin by defining the trace and helicity contractions 
\begin{align}\label{helicity_contractions}
&T(\q) = \delta_{ij} T_{ij}(\q), \qquad
T^{(s)}(\q)= \half\ep^{(s)}_{ij}(-\q) T_{ij}(\q), \qquad \Upsilon(\q_1,\q_2) =\delta_{ij}\delta_{kl} \Upsilon_{ijkl}(\q_1,\q_2),\\ 
& \Upsilon^{(s_2)}(\q_1,\q_2) =\half \delta_{ij}\ep^{(s_2)}_{kl}(-\q_2) \Upsilon_{ijkl}(\q_1,\q_2), \quad
\Upsilon^{(s_1 s_2)}(\q_1,\q_2) =\frac{1}{4} \ep^{(s_1)}_{ij}(-\q_1) \ep^{(s_2)}_{kl}(-\q_2) \Upsilon_{ijkl}(\q_1,\q_2).\nn
\end{align} 
Here, the transverse traceless polarisation tensors $\ep_{ij}^{(s)}(\q)$ carry a helicity index $s=\pm 1$ (see appendix \ref{helicity_basis_app} for a summary of our conventions). 
To write our results in compact form, it is also useful to introduce the general decomposition
\[\label{ABdef}
\ll T_{ij}(\q)T_{kl}(-\q)\rr = A(q)\Pi_{ijkl}+B(q)\pi_{ij}\pi_{kl},
\]
where the transverse traceless and transverse projection operators, $\Pi_{ijkl}$ and $\pi_{ij}$ respectively, are defined by 
\[\label{projection_operators}
\Pi_{ijkl}=\frac{1}{2}(\pi_{ik}\pi_{jl}+\pi_{il}\pi_{jk}-\pi_{ij}\pi_{kl}),\qquad
\pi_{ij}=\delta_{ij}-\frac{q_iq_j}{q^2}.
\]
The form of this decomposition follows directly from the Ward identity \eqref{2ptWard} in momentum space.  Physically, $A(q)$ encodes the transverse traceless and $B(q)$ the trace part of the stress tensor 2-point function.  Moreover, from \eqref{2ptTaylor}, in the squeezed limit as $q_1\tto 0$ we have
\[\label{ABsq}
A(q_1) = O(q_1^2), \qquad B(q_1) = O(q_1^2).
\]

With these considerations in place, our aim is now to project the Taylor expansion \eqref{Taylor} for the squeezed limit of the 3-point function into the helicity basis.
To commute polarisation tensors inside momentum derivatives where needed we use the additional identity
\[\label{ep_deriv}
\frac{\p}{\p q_a}\ep_{ij}^{(s)}(\q)=-\frac{2}{q^2}q_{(i}\ep_{j)a}^{(s)}(\q),
\]
as derived in Appendix \ref{momderiv}.
From this relation (or directly from \eqref{projection_operators}), we also have
\[\label{pi_deriv}
\frac{\p}{\p q_a}\pi_{ij} = -\frac{2}{q^2}q_{(i}\pi_{j)a}, \qquad 
\frac{\p}{\p q_a}\Pi_{ijkl} = -\frac{2}{q^2}\Big(q_{(i}\Pi_{j)akl}+q_{(k}\Pi_{l)aij}\Big).
\]  
To evaluate the various contractions of polarisation tensors arising on the right-hand sides, without loss of generality we can simply pick a basis where $\q$ and $\q_1$ lie in the $(x,z)$ plane and use the explicit representation given in appendix \ref{helicity_app}.
This allows all contractions of polarisation tensors to be evaluated in terms of the momentum magnitudes $q$ and $q_1$ and the angle $\vphi$ defined in \eqref{vphi_def}.

After some straightforward computation, we then find
%{\allowdisplaybreaks
\begin{align}
\label{QFTsoft1}
& \ll T(q_1)T(q_2)T(q_3)\rr %\nn\\[1ex]&\quad
-2\ll T(q_1)\Y(q_2,q_3)\rr
- 2\ll \Y(q_1,q_2)T(q_3)\rr - 2\ll\Y(q_1,q_3)T(q_2)\rr 
\nn\\[1ex]&\quad 
=4\big(B(q)-q B'(q)\big)+O(q_1^2),\\[2ex]
%%%%%%%%%%%%%%%%%%%%%%%%
\label{QFTsoft2}
&\ll T(q_1)T(q_2)T^{(s_3)}(q_3)\rr\nn\\[1ex]&\quad 
-2\ll T(q_1)\Y^{(s_3)}(q_2,q_3)\rr
- 2\ll \Y(q_1,q_2)T^{(s_3)}(q_3)\rr-2\ll\Y^{(s_3)}(q_1,q_3)T(q_2)\rr
\nn\\[1ex]&\quad
= O(q_1^2),\\[2ex]
%%%%%%%%%%%%%%%%%%%%%%%%
\label{QFTsoft3}
& \ll T(q_1)T^{(s_2)}(q_2)T^{(s_3)}(q_3)\rr \nn\\[1ex]&\quad
-2\ll T(q_1)\Y^{(s_2s_3)}(q_2,q_3)\rr
- 2\ll \Y^{(s_2)}(q_1,q_2)T^{(s_3)}(q_3)\rr 
- 2\ll\Y^{(s_3)}(q_1,q_3)T^{(s_2)}(q_2)\rr 
\nn\\[1ex]&\quad 
= \frac{1}{2}\big(7A(q)-qA'(q)\big)\delta^{s_2s_3}+O(q_1^2), %\\[2ex]
%%%%%%%%%%%%%%%%%%%%%%
\end{align}
\begin{align}
\label{QFTsoft4}
&\ll T^{(s_1)}(q_1)T(q_2)T(q_3)\rr\nn\\[1ex]&\quad
-2\ll T^{(s_1)}(q_1)\Y(q_2,q_3)\rr 
-2\ll\Y^{(s_1)}(q_2,q_1)T(q_3)\rr-2\ll \Y^{(s_1)}(q_3,q_1)T(q_2)\rr\nn\\[1ex]&\quad
=-\sqrt{2}\sin^2\vphi\, \big(2B(q)+qB'(q)\big)+O(q_1^2),\\[2ex]
%%%%%%%%%%%%%%%%%%%%%%%
\label{QFTsoft5}
&\ll T^{(s_1)}(q_1)T^{(s_2)}(q_2)T(q_3)\rr \nn\\[1ex]&\quad
-2\ll T^{(s_1)}(q_1)\Y^{(s_2)}(q_3,q_2)\rr 
-2\ll\Y^{(s_1)}(q_3,q_1)T^{(s_2)}(q_2)\rr
-2\ll\Y^{(s_1 s_2)}(q_1,q_2)T(q_3)\rr 
\nn\\[1ex]&\quad
=\frac{1}{4}(\cos\vphi-s_1s_2)^2\big(A(q)+2B(q)\big)
+\frac{q_1}{4q}A(q)\sin^2\vphi(s_1 s_2-\cos\vphi)\nn\\[1ex]&\qquad
+ \frac{q_1}{8q}\Big[q\big(2B'(q)-A'(q)\big)+2B(q)\Big]
\cos\vphi(s_1 s_2-\cos\vphi)^2 
+O(q_1^2), \\[2ex]
%%%%%%%%%%%%%%%%%%%%%%%%%%%
\label{QFTsoft6}
&\ll T^{(s_1)}(q_1)T^{(s_2)}(q_2)T^{(s_3)}(q_3)\rr \nn\\[1ex]&\quad
-2\ll T^{(s_1)}(q_1)\Y^{(s_2s_3)}(q_2,q_3)\rr
-2\ll\Y^{(s_1s_2)}(q_1,q_2)T^{(s_3)}(q_3)\rr
-2\ll\Y^{(s_1s_3)}(q_1,q_3)T^{(s_2)}(q_2)\rr 
\nn\\[1ex]&\quad
= -\frac{1}{4\sqrt{2}}\sin^2\vphi\,\Big[\big(qA'(q) +2A(q)\big)\delta^{s_2s_3} 
+ \frac{q_1}{q}\big(q A'(q)+A(q)\big)s_1(s_2+s_3)\Big]
+O(q_1^2),
\end{align}
%}
where $A'(q) = \d A(q)/\d q$, etc., and the squeezed limit $q_1\tto 0$ is taken while enforcing \eqref{q2q3}.  In deriving \eqref{QFTsoft4} and \eqref{QFTsoft5} we also swapped the indices on $\Y_{klmn}$ around using 
\eqref{Yswap}.

The soft theorems \eqref{QFTsoft1}-\eqref{QFTsoft6}
are the main result of this paper from a quantum field theory perspective.
For free fields, we have explicitly checked each of
these relations using the results of \cite{Bzowski:2011ab, McFadden:2010vh}, in which the correlators above were evaluated for minimal and conformal scalars, fermions and gauge fields.  
Given all the different field types and polarisations, this amounts to a large number of non-trivial checks.  To quote just a single example, for a minimal scalar we have
{\allowdisplaybreaks
\begin{align}
& \ll T(q_1)T(q_2)T(q_3) \rr = \frac{1}{128}\Big( 2q_1q_2q_3-(q_1+q_2+q_3)(q_1^2+q_2^2+q_3^2)\Big),\\[1ex]
&\ll \Y(q_1,q_2)T(q_3)\rr = 0, \qquad \ll T(q)T(-q)\rr = 4B(q)=\frac{1}{64}q^3.
\end{align}
}
Imposing \eqref{q2q3}, in the squeezed limit we then recover 
\begin{align}
&\ll T(q_1)T(q_2)T(q_3)\rr
-2\ll T(q_1)\Y(q_2,q_3)\rr
- 2\ll \Y(q_1,q_2)T(q_3)\rr - 2\ll\Y(q_1,q_3)T(q_2)\rr \nn\\&
= -\frac{1}{32}q^3+O(q_1^2),
\end{align}
precisely as predicted.

\section{Soft limits in holographic cosmology}\label{sec_hc}

Armed with our results \eqref{QFTsoft1}-\eqref{QFTsoft6} for the soft limit of the stress tensor 3-point function, we are now in a position to derive the cosmological consistency relations holographically.  After introducing the necessary holographic formulae for cosmological correlators in section \ref{hol_form_sec}, we deduce the consistency relations in section \ref{consistency_sec}.

\subsection{Holographic formulae}\label{hol_form_sec}

In holographic cosmology, the 2-point functions of superhorizon cosmological curvature perturbations $\z(q)$ and gravitons $\g^{(s)}(q)$ are given by \cite{McFadden:2009fg} 
\[
\label{2ptformulae}
 \<\!\<\z(q)\z(-q)\>\!\> = \frac{-1}{8\Im[B(q)]}, \qquad \<\!\<\bg^{(s)}(q)\bg^{(s')}(-q)\>\!\> = \frac{-\delta^{ss'}}{\Im[A(q)]},
\]
where $A(q)$ and $B(q)$ are the transverse traceless and trace pieces of the stress tensor 2-point function in the dual QFT, as defined in \eqref{ABdef}.
The imaginary part in these formulae is taken after making the analytic continuation 
\[ \label{an_cont}
N \tto -iN, \qquad q \tto -iq,
\]
where $N$ is  the rank of the gauge group of the dual QFT.  This continuation is the dual QFT analogue of the bulk analytic continuation effecting the domain-wall/cosmology correspondence, which acts to map perturbations on a domain-wall background to perturbations on a corresponding cosmological background.  For a detailed explanation of our approach to holographic cosmology (and of how the rank $N$ appears in the correlators of the dual QFT) we refer the reader to \cite{McFadden:2009fg, McFadden:2010na, McFadden:2010vh, McFadden:2011kk, Bzowski:2011ab}.  For our present purposes, however, no further information is necessary.  
In fact, we do not even need to know how $N$ enters any of the correlators: to derive the cosmological consistency relations we need only to relate 3-point functions in the squeezed limit to 2-point functions, and as we will see, this does not require knowing the $N$-dependence of correlators.

The holographic formulae relating
3-point cosmological correlators to correlators of the dual QFT were derived in \cite{McFadden:2010vh, McFadden:2011kk}, and read 
%{\allowdisplaybreaks
\begin{align}
\label{holo_zzz}
&\<\!\<\z(q_1)\z(q_2)\z(q_3)\>\!\>  \nn\\[1ex]&\quad
 = -\frac{1}{256}\Big(\prod_i \Im [B(q_i)]\Big)^{-1}\times
\Im \Big[\<\!\<T(q_1)T(q_2)T(q_3)\>\!\> + 4\sum_i B(q_i) \nn\\[0ex]&\qquad\qquad\qquad\qquad\qquad\qquad\qquad\qquad
-2\Big( \<\!\<T(q_1)\Upsilon(q_2,q_3)\>\!\>+\mathrm{cyclic\,perms.}\Big)\Big], \\[2ex]
%\\[-2ex]
%%%%%%%%%%%%%%%%%%%%
\label{holo_zzg}
&\<\!\<\z(q_1)\z(q_2)\bg^{(s_3)}(q_3)\>\!\>  \nn\\[1ex]& \quad
= -\frac{1}{32} \Big(\Im[B(q_1)]\Im[B(q_2)]\Im[A(q_3)]\Big)^{-1} \nn\\[1ex]& \qquad
\times \Im\Big[\<\!\<T(q_1)T(q_2)T^{(s_3)}(q_3)\>\!\> 
-2\big(\Theta_1^{(s_3)}B(q_1)+\Theta_2^{(s_3)}B(q_2)\big)  \nn\\[1ex]&\qquad\qquad\quad
-2\<\!\<\Upsilon(q_1,q_2)T^{(s_3)}(q_3)\>\!\>
-2\<\!\<T(q_1)\Upsilon^{(s_3)}(q_2,q_3)\>\!\>  
-2\<\!\<T(q_2)\Upsilon^{(s_3)}(q_1,q_3)\>\!\>
\Big],  %\\[2ex]
%\\[0ex]
%%%%%%%%%%%%%%%%%%%
\end{align}
\begin{align}
\label{holo_zgg}
& \<\!\<\z(q_1)\bg^{(s_2)}(q_2)\bg^{(s_3)}(q_3)\>\!\> \nn\\[1ex]&\quad
= -\frac{1}{4}\Big(\Im[B(q_1)]\Im[A(q_2)]\Im[A(q_3)]\Big)^{-1} \nn\\[1ex]&\qquad
\times \Im\Big[\<\!\<T(q_1)T^{(s_2)}(q_2)T^{(s_3)}(q_3)\>\!\>
-\half \big(A(q_2)+A(q_3)\big)\theta^{(s_2s_3)} 
-B(q_1) \Theta^{(s_2s_3)}  \nn\\[1ex]&\qquad\quad
-2\<\!\<T(q_1)\Upsilon^{(s_2s_3)}(q_2,q_3)\>\!\> 
-2\<\!\<T^{(s_2)}(q_2)\Upsilon^{(s_3)}(q_1,q_3)\>\!\>
-2\<\!\<T^{(s_3)}(q_3)\Upsilon^{(s_2)}(q_1,q_2)\>\!\>
\Big], \\[2ex]
%\\[0ex]
%%%%%%%%%%%%%%%%%%%%
\label{holo_ggg}
&\<\!\<\bg^{(s_1)}(q_1)\bg^{(s_2)}(q_2)\bg^{(s_3)}(q_3)\>\!\> \nn\\[1ex]&\quad
= -\Big(\prod_i\Im[A(q_i)]\Big)^{-1}  
\times\Im \Big[2\<\!\<T^{(s_1)}(q_1)T^{(s_2)}(q_2)T^{(s_3)}(q_3)\>\!\> 
-\frac{1}{2}\Theta^{(s_1s_2s_3)}\sum_i A(q_i) \nn\\[1ex]&\qquad \qquad\qquad\qquad \qquad\qquad\qquad
-4\Big(\<\!\<T^{(s_1)}(q_1)\Upsilon^{(s_2s_3)}(q_2,q_3)\>\!\> + \mathrm{cyclic\,perms.}\Big)
 \Big]. 
\end{align}
%}
Just as in the dual QFT, the double bracket notation used here for cosmological correlators simply indicates the removal of  the overall momentum-conserving delta function, e.g.,
\begin{align}
\<\z(\q_1)\z(\q_2)\z(\q_3)\> = \<\!\<\z(q_1)\z(q_2)\z(q_3)\>\!\>(2\pi)^3\delta(\q_1+\q_2+\q_3).
\end{align}
The quantities $\Theta_1^{(s)}$, $\Theta_2^{(s)}$,
$\Theta^{(s_1 s_2)}$, $\theta^{(s_1 s_2)}$ and $\Theta^{(s_1 s_2 s_3)}$
represent specific contractions of polarisation tensors and projection operators and are listed in appendix \ref{helicity_app}. The imaginary parts in these formulae are taken after applying the continuation
\eqref{an_cont}, as above.
We emphasize also the presence of the semilocal contact terms in the numerators of these formulae (i.e., the terms non-analytic in only a single momenta).  The form of these terms was carefully derived in \cite{McFadden:2010vh, McFadden:2011kk}; to obtain the correct cosmological consistency relations it is essential their structure is correct.

\subsection{Cosmological consistency relations}\label{consistency_sec}

To derive the cosmological consistency relations we simply need to insert our results \eqref{QFTsoft1}-\eqref{QFTsoft6} for the soft limit of the stress tensor 3-point function (along with \eqref{ABsq} for the 2-point function) into the holographic formulae \eqref{holo_zzz}-\eqref{holo_ggg}.  The resulting terms involving the 2-point function of the stress tensor and its derivatives can then be replaced with cosmological 2-point functions and their derivatives using \eqref{2ptformulae}.  The helicity contraction terms (i.e., $\Theta_1^{(s)}$, $\Theta_2^{(s)}$,
$\Theta^{(s_1 s_2)}$, $\theta^{(s_1 s_2)}$ and $\Theta^{(s_1 s_2 s_3)}$) are all dimensionless functions of the momentum magnitudes (see appendix \ref{helicity_app}) and so do not contribute to the imaginary part of any formulae.
Similarly, the momentum derivatives in \eqref{QFTsoft1}-\eqref{QFTsoft6} all appear in dimensionless combinations and 
do not contribute to the imaginary part either.
Knowing the dependence of correlators on the rank $N$ of the QFT gauge group is not necessary:
the QFT soft theorems giving rise to \eqref{QFTsoft1}-\eqref{QFTsoft6} are quite independent of this, and the same analytic continuation applies in all holographic formulae, so the imaginary part of the stress tensor 3-point function is simply related to the imaginary part of the stress tensor 2-point function, and hence to the cosmological 2-point function. 

The complete leading and sub-leading behaviour of cosmological 3-point functions in the squeezed limit can therefore be written
{\allowdisplaybreaks
\begin{align}
\label{cosmosoft1}
\frac{ \ll \z(q_1)\z(q_2)\z(q_3)\rr}{\ll\z(q_1)\z(-q_1)\rr} &=
(1-n_S(q))\ll\z(q)\z(-q)\rr + O(q_1^2), \\[2ex]
%%%%%%%%%%%%%%%%
\label{cosmosoft2}
\frac{\ll\z(q_1)\z(q_2)\g^{(s_3)}(q_3)\rr}{\ll\z(q_1)\z(-q_1)\rr} &= O(q_1^2),\\[2ex]
%%%%%%%%%%%%%%%%%
\label{cosmosoft3}
\frac{\ll \z(q_1)\g^{(s_2)}(q_2)\g^{(s_3)}(q_3)\rr}{\ll\z(q_1)\z(-q_1)\rr} &=
-n_T(q)\ll\g^{(+)}(q)\g^{(+)}(-q)\rr\delta^{s_2s_3} + O(q_1^2), \\[2ex]
%%%%%%%%%%%%%%%%%%%%
\label{cosmosoft4}
\frac{\ll\g^{(s_1)}(q_1)\z(q_2)\z(q_3)\rr}{\ll\g^{(+)}(q_1)\g^{(+)}(-q_1)\rr}&=
(4-n_S(q))\ll\z(q)\z(-q)\rr \frac{1}{2\sqrt{2}}\sin^2\vphi + O(q_1^2), \\[2ex]
%%%%%%%%%%%%%%%%%%%%%
\label{cosmosoft5}
\frac{\ll\g^{(s_1)}(q_1)\g^{(s_2)}(q_2)\z(q_3)\rr}{\ll\g^{(+)}(q_1)\g^{(+)}(-q_1)\rr} &= 
-\frac{1}{16}\frac{q_1}{q} \ll\g^{(+)}(q)\g^{(+)}(-q)\rr\cos\vphi(s_1s_2-\cos\vphi)^2\nn\\[1ex]&\qquad
 +O(q_1^2), \\[2ex]
%%%%%%%%%%%%%%%%%%%%
\label{cosmosoft6}
\frac{\ll\g^{(s_1)}(q_1)\g^{(s_2)}(q_2)\g^{(s_3)}(q_3)\rr}{\ll\g^{(+)}(q_1)\g^{(+)}(-q_1)\rr}&=
(3-n_T(q))\ll\g^{(+)}(q)\g^{(+)}(-q)\rr 
\frac{1}{2\sqrt{2}}\sin^2\vphi \nn\\[1ex] &\qquad
\times \Big(\delta^{s_2s_3} 
+s_1(s_2+s_3)\frac{q_1}{q}\Big)+O(q_1^2),
\end{align}
}
where the limit $q_1\tto 0$ is taken while imposing \eqref{q2q3}.
The scalar and tensor tilts in these formulae are defined by the usual expressions,
\[
n_S(q)-1 = \frac{\d}{\d\ln q}\ln\Delta_S^2(q), \qquad
n_T(q) = \frac{\d}{\d\ln q}\ln\Delta_T^2(q),
\]
where the corresponding power spectra are 
\[
\Delta_S^2(q) = \frac{q^3}{2\pi^2}\ll\z(q)\z(-q)\rr,\qquad
\Delta_T^2(q) = \frac{2q^3}{\pi^2}\ll\g^{(+)}(q)\g^{(+)}(-q)\rr.
\]
(As per convention, a scale-invariant spectrum thus corresponds to $n_S=1$ but $n_T=0$.)  In writing the results in the form above we have also assumed invariance under parity, which acts to invert all graviton helicities.

Our holographic derivation of the cosmological squeezed limits \eqref{cosmosoft1}-\eqref{cosmosoft6} is the main result of this paper.  The leading term of \eqref{cosmosoft1} is the famous Maldacena consistency relation for the scalar bispectrum \cite{Maldacena:2002vr}, while the absence of any contribution at order $q_1$ is equivalent to the `conformal' consistency relation of \cite{Creminelli:2011rh, Creminelli:2012ed}.  The leading pieces of \eqref{cosmosoft3} and \eqref{cosmosoft4} were 
likewise discovered in \cite{Maldacena:2002vr}, while the subleading piece of \eqref{cosmosoft4} is equivalent to the relation found in \cite{Creminelli:2012ed}.\footnote{Note that the example (68) given in \cite{Creminelli:2012ed} has a linear term because they set $\vec{k}_2=-\vec{k}_1-\q$, instead of the symmetric definition we use here  \eqref{q2q3}.}  The remaining relations are then equivalent to the $n=0,1$ relations of \cite{Hinterbichler:2013dpa}.

\section{Explicit tests}\label{sec_tests}

As a check on our calculations, we now verify the cosmological consistency relations \eqref{cosmosoft1}-\eqref{cosmosoft6} in two distinct scenarios, standard single field slow-roll inflation and strongly coupled holographic cosmologies based on a perturbative dual QFT.

\subsection{Slow-roll inflation}

%We now turn to verify our predicted soft limits explicitly in the case of standard single-field slow-roll inflation.  
While the consistency relations for slow-roll inflation have been checked elsewhere (see, e.g., \cite{Maldacena:2002vr, Creminelli:2011rh, Creminelli:2012ed, Sreenath:2014nka}), 
we repeat the exercise here since, relative to the literature, we have chosen to replace contractions of polarisation tensors with explicit expressions in terms of $q$, $q_1$ and $\vphi$, and in addition our symmetrised definition \eqref{q2q3} of the squeezed limit was not always applied.

The 2-point functions at leading order in slow-roll are \cite{WeinbergBook}
\[
\ll \z(q)\z(-q)\rr_{SR} = \frac{\kappa^2 H_*^2}{4 \ep_* q^3}, \qquad
\ll \bg^{(s)}(q)\bg^{(s')}(-q)\rr_{SR} = \frac{\kappa^2 H_*^2}{q^3}\delta^{s s'},
\]
where $\K^2= 8\pi G$ and the \textit{SR} on correlators is shorthand for `slow-roll'.  The asterisk indicates as usual the evaluation of quantities at the moment of horizon crossing for which $q=aH$. Our slow-roll parameters are defined as
\[
\ep = -\frac{\dot{H}}{H^2}, \qquad \eta = \frac{\ddot{\phi}}{\dot{\phi}H},
\]
where $H = \dot{a}/a$ is the proper Hubble rate and $\phi$ is the inflaton.  With this choice, the spectral tilts are then
\[
n_S -1 = -4\ep_*-2\eta_*,\qquad n_T = -2\ep_*,
\]
to leading order in slow-roll.

The 3-point functions for slow-roll inflation were obtained in \cite{Maldacena:2002vr}.  Replacing the contractions of polarisation tensors that appear in \cite{Maldacena:2002vr} with expressions involving the momentum magnitudes alone, we find \cite{Bzowski:2011ab}
%{\allowdisplaybreaks
\begin{align}
\<\!\<\z(q_1)\z(q_2)\z(q_3)\>\!\>_{SR} &=
\frac{\K^4 H_*^4}{32 \ep_*^2}\frac{1}{c_{123}^3}\Big[2\eta_* \big(a_{123}^3-3a_{123}b_{123}+3c_{123}\big) \nn\\[1ex]&\qquad\qquad
+\ep_*\Big(a_{123}^3-2a_{123}b_{123}-16c_{123}+\frac{8 b_{123}^2}{a_{123}}\Big)\Big],
\\[1ex]
%%%%%%%%%%%%%%%%%%%%
\<\!\<\z(q_1)\z(q_2)\bg^{(+)}(q_3)\>\!\>_{SR} & = \frac{\K^4 H_*^4}{16\sqrt{2}\ep_*} \frac{\lambda^2}{a_{123}^2 c_{123}^3 q_3^2} \big[a_{123}^3-a_{123}b_{123}-c_{123}\big], %\\[2ex]
\end{align}
\begin{align}
%%%%%%%%%%%%%%%
\<\!\<\z(q_1)\bg^{(+)}(q_2)\bg^{(+)}(q_3)\>\!\>_{SR} &= -\frac{\K^4H_*^4}{128 b_{23}^5 q_1^2}(q_1^2-a_{23}^2)^2
\Big[(q_1^2-a_{23}^2+2b_{23})-\frac{8 b_{23}^2}{q_1 a_{123}}\Big],
\\[2ex]
%%%%%%%%%%%%%%
\<\!\<\z(q_1)\bg^{(+)}(q_2)\bg^{(-)}(q_3)\>\!\>_{SR} &= -\frac{\K^4H_*^4}{128\, b_{23}^5 q_1^2}(q_1^2-a_{23}^2+4b_{23})^2\nn\\[-1ex]&\qquad\qquad\qquad\qquad
\times\Big[(q_1^2-a_{23}^2+2b_{23})-\frac{8 b_{23}^2}{q_1 a_{123}}\Big],
\\[1ex]
%\label{SRgggbispec}
%%%%%%%%%%%%%%%%%%%%%%%%
\<\!\<\bg^{(+)}(q_1)\bg^{(+)}(q_2)\bg^{(+)}(q_3)\>\!\>_{SR} &= \frac{\K^4H_*^4}{64 \sqrt{2}}\,\frac{\lambda^2a_{123}^2}{c_{123}^5}(a_{123}^3-a_{123}b_{123}-c_{123}),  \\[1ex]
%%%%%%%%%%%%%%%%%%%%%%%%
 \<\!\<\bg^{(+)}(q_1)\bg^{(+)}(q_2)\bg^{(-)}(q_3)\>\!\>_{SR} &=
\frac{\K^4H_*^4}{64\sqrt{2}}\,\frac{\lambda^2}{a_{123}^2c_{123}^5}(q_3-a_{12})^4(a_{123}^3-a_{123}b_{123}-c_{123}),
\end{align}
%}
where we use the following shorthand notation for symmetric polynomials of the momentum magnitudes
\begin{align}\label{abc_def}
& a_{123} =q_1+q_2+q_3,\qquad  b_{123}  = q_1q_2+q_2q_3+q_3q_1,\qquad c_{123} = q_1q_2q_3 \nn\\
& a_{12}=q_1+q_2,\quad\quad\quad\quad\quad b_{12} = q_1q_2,
\end{align}
and similarly for $a_{23}$ and $b_{23}$, etc.  We also define the useful combination 
\[\label{lambda_def}
\lambda^2 = (q_1+q_2+q_3)(-q_1+q_2+q_3)(q_1-q_2+q_3)(q_1+q_2-q_3) =
-a_{123}(a_{123}^3-4a_{123}b_{123}+8c_{123}).
\]
By Heron's formula, $\lambda$ is a quarter the area of the triangle with side lengths $q_1$, $q_2$ and $q_3$.

To obtain the behaviour in the squeezed limit, we first set
\[\label{q2q3_magnitudes}
q_2^2 = q^2-q_1 q \cos\vphi+\frac{1}{4}q_1^2,\qquad
q_3^2 = q^2+q_1 q \cos\vphi+\frac{1}{4}q_1^2,
\]
as per \eqref{q2q3}, and then by direct evaluation we find
{\allowdisplaybreaks
\begin{align}
\Big[\frac{ \ll \z(q_1)\z(q_2)\z(q_3)\rr}{\ll\z(q_1)\z(-q_1)\rr}\Big]_{SR} &=
(2\ep_*+\eta_*)\,\frac{\K^2 H_*^2}{2\ep_* q^3}
+ O(q_1^2), \\[2ex]
%%%%%%%%%%%%%%%%
\Big[\frac{\ll\z(q_1)\z(q_2)\g^{(s_3)}(q_3)\rr}{\ll\z(q_1)\z(-q_1)\rr}\Big]_{SR} &= O(q_1^2),\\[2ex]
%%%%%%%%%%%%%%%%%
\Big[\frac{\ll \z(q_1)\g^{(s_2)}(q_2)\g^{(s_3)}(q_3)\rr}{\ll\z(q_1)\z(-q_1)\rr}\Big]_{SR} &=
2\ep_*\frac{\K^2 H_*^2}{q^3}\delta^{s_2s_3} + O(q_1^2), \\[2ex]
%%%%%%%%%%%%%%%%%%%%
\label{SR_soft_gzz}
\Big[\frac{\ll\g^{(s_1)}(q_1)\z(q_2)\z(q_3)\rr}{\ll\g^{(+)}(q_1)\g^{(+)}(-q_1)\rr}\Big]_{SR} &=
\frac{\K^2 H_*^2}{\ep_* q^3} \frac{3}{8\sqrt{2}}\sin^2\vphi + O(q_1^2), \\[2ex]
%%%%%%%%%%%%%%%%%%%%%
\Big[\frac{\ll\g^{(s_1)}(q_1)\g^{(s_2)}(q_2)\z(q_3)\rr}{\ll\g^{(+)}(q_1)\g^{(+)}(-q_1)\rr}\Big]_{SR} &= 
-\frac{\K^2H_*^2}{16 \,q^4}\,q_1\cos\vphi(s_1s_2-\cos\vphi)^2
 +O(q_1^2), \\[2ex]
%%%%%%%%%%%%%%%%%%%%
\label{SR_soft_ggg}
\Big[\frac{\ll\g^{(s_1)}(q_1)\g^{(s_2)}(q_2)\g^{(s_3)}(q_3)\rr}{\ll\g^{(+)}(q_1)\g^{(+)}(-q_1)\rr}\Big]_{SR} &=
\frac{\K^2H_*^2}{q^3} 
\frac{3}{2\sqrt{2}}\sin^2\vphi \Big(\delta^{s_2s_3} 
+s_1(s_2+s_3)\frac{q_1}{q}\Big)+O(q_1^2).
\end{align}
}
These squeezed limits are indeed precisely in accordance with our consistency relations \eqref{cosmosoft1}-\eqref{cosmosoft6}.  In comparing \eqref{SR_soft_gzz} and \eqref{SR_soft_ggg} with \eqref{cosmosoft4} and \eqref{cosmosoft6} respectively, note that it is sufficient to take $(4-n_S) = 3+O(\ep_*,\eta_*)$ and $(3-n_T) = 3+O(\ep_*)$, since we have only evaluated the left-hand sides to leading order in slow-roll.

\subsection{Strongly coupled holographic cosmologies}

We now turn to verify the consistency relations for strongly coupled holographic cosmologies based on a perturbative dual QFT.  For a detailed discussion of these cosmologies, including their predictions and their fit to recent observational data, we refer the reader to \cite{McFadden:2009fg, McFadden:2010na, McFadden:2010vh, McFadden:2010jw, Easther:2011wh, Bzowski:2011ab}.
In short, we postulate a phenomenological dual QFT consisting of three-dimensional $SU(N)$ Yang-Mills theory with general adjoint matter and interactions, then use the holographic formulae to extract cosmological predictions for the regime in which the dual QFT is weakly coupled.  
The basic parameters of the model are thus the rank $N$, the Yang-Mills coupling $\gYM^2$ (or more accurately, the dimensionless effective coupling $\geff^2=\gYM^2 N/q$, which is assumed to be small  over the range of scales relevant to the CMB), and the field content, i.e., the number of minimal and conformal scalars, fermions, and gauge fields.  In fact, at leading 1-loop order the interactions do not contribute and $\geff^2$ does not appear, although at 2-loop order interactions generate deviations from scale invariance of the form $n_S(q){-}1\sim \geff^2$ and $n_T(q)\sim \geff^2$ \cite{McFadden:2009fg, McFadden:2010na}. Moreover, the field content only enters (with one exception) in two specific combinations $\mathcal{N}_{(A)}$ and $\mathcal{N}_{(B)}$ defined in (4.6) of \cite{Bzowski:2011ab}; the former effectively counts the total number of fields, while the latter counts the number of non-conformal fields (i.e., minimal scalars plus gauge fields).

Evaluating the leading 1-loop Feynman diagrams contributing to the 2- and 3-point functions of the stress tensor then using the holographic formulae \eqref{2ptformulae} and \eqref{holo_zzz}-\eqref{holo_ggg}, one obtains cosmological predictions as follows.
Firstly, the 2-point functions read \cite{McFadden:2009fg}
\[
\ll \z(q)\z(-q)\rr_{HM} = \frac{32}{N^2 \mathcal{N}_{(B)} q^3}, \qquad
\ll \bg^{(s)}(q)\bg^{(s')}(-q)\rr_{HM} = \frac{256}{N^2 \mathcal{N}_{(A)}q^3}\delta^{s s'},
\]
where the subscript \textit{HM} stands for `holographic model'. At 1-loop order, the spectrum is thus scale-invariant (i.e., $n_S(q) = 1$, $n_T(q)=0$), but this does not persist at higher orders as noted above.  We observe also that the large-$N$ 't Hooft limit of the dual QFT is consistent with the small observed amplitude of the scalar power spectrum.

Next, from \cite{McFadden:2010vh, Bzowski:2011ab}, the 3-point functions are
%{\allowdisplaybreaks
\begin{align}
\label{main_hol_results}
&\<\!\<\z(q_1)\z(q_2)\z(q_3)\>\!\>_{HM}
=  \frac{512}{N^4\mathcal{N}_{(B)}^2}\frac{\lambda^2}{a_{123}c_{123}^3},
%\\[2ex]
\end{align}
\begin{align}
%%%%%%%%%%%%%
&\<\!\<\z(q_1)\z(q_2)\bg^{(+)}(q_3)\>\!\>_{HM}  = \frac{2048}{\sqrt{2}N^4 \mathcal{N}_{(A)}\mathcal{N}_{(B)}} \frac{\lambda^2}{a_{123}^2 c_{123}^3 q_3^2} %\nn\\[1ex]&\qquad\qquad\times 
\Big[a_{123}^3-a_{123}b_{123}-c_{123}-a_{123}q_3^2\Big],\\[1ex]
%%%%%%%%%%%%
&\<\!\<\z(q_1)\bg^{(+)}(q_2)\bg^{(+)}(q_3)\>\!\>_{HM} = -\frac{512}{N^4\mathcal{N}_{(A)}^2  b_{23}^5 q_1^2}(q_1^2-a_{23}^2)^2
\Big[q_1^2-a_{23}^2+2b_{23}+\frac{32 b_{23}^3}{a_{123}^4}\Big],
\\[2ex]
%%%%%%%%%%%%%
&\<\!\<\z(q_1)\bg^{(+)}(q_2)\bg^{(-)}(q_3)\>\!\>_{HM} = -\frac{512}{N^4\mathcal{N}_{(A)}^2 b_{23}^5 q_1^2}(q_1^2-a_{23}^2+4b_{23})^2(q_1^2-a_{23}^2+2b_{23}),
\\[2ex]
%%%%%%%%%%%%%
& \<\!\<\bg^{(+)}(q_1)\bg^{(+)}(q_2)\bg^{(+)}(q_3)\>\!\>_{HM}
=\frac{1024}{\sqrt{2}N^4 \mathcal{N}_{(A)}^2}\frac{\lambda^2 a_{123}^2}{c_{123}^5} \Big[ a_{123}^3-a_{123}b_{123}-c_{123}
\nn\\[-1ex]&\qquad\qquad\qquad\qquad\qquad\qquad\qquad\qquad
\qquad\qquad\qquad\qquad
-\Big(1-4\frac{\mathcal{N}_\psi}{\mathcal{N}_{(A)}}\Big)\frac{64 c_{123}^3}{a_{123}^6}\Big], \\[1ex]
%%%%%%%%%%%%%
& \<\!\<\bg^{(+)}(q_1)\bg^{(+)}(q_2)\bg^{(-)}(q_3)\>\!\>_{HM} = \frac{1024}{\sqrt{2}N^4 \mathcal{N}_{(A)}^2}\frac{\lambda^2}{a_{123}^2c_{123}^5}(q_3-a_{12})^4(a_{123}^3-a_{123}b_{123}-c_{123}),
\end{align}
%}
where the symmetric polynomials $a_{123}$, $a_{23}$, $\lambda^2$, etc., are as defined earlier in \eqref{abc_def} and \eqref{lambda_def}.  The $\mathcal{N}_{\psi}$ is the penultimate formula is the number of fermions in the dual QFT (the one exception where the field content does not enter as just $\mathcal{N}_{(A)}$ or $\mathcal{N}_{(B)}$ as noted above).

To extract the behaviour of the holographic model in the squeezed limit, we once again set $q_2$ and $q_3$ as in \eqref{q2q3_magnitudes}, then by direct evaluation we find
{\allowdisplaybreaks
\begin{align}
\Big[\frac{ \ll \z(q_1)\z(q_2)\z(q_3)\rr}{\ll\z(q_1)\z(-q_1)\rr}\Big]_{HM} &= O(q_1^2),\\[2ex]
%%%%%%%%%%%%%%%%
\Big[\frac{\ll\z(q_1)\z(q_2)\g^{(s_3)}(q_3)\rr}{\ll\z(q_1)\z(-q_1)\rr}\Big]_{HM} &= O(q_1^2),\\[2ex]
%%%%%%%%%%%%%%%%%
\Big[\frac{\ll \z(q_1)\g^{(s_2)}(q_2)\g^{(s_3)}(q_3)\rr}{\ll\z(q_1)\z(-q_1)\rr}\Big]_{HM} &=O(q_1^2),\\[2ex]
%%%%%%%%%%%%%%%%%%%%
\Big[\frac{\ll\g^{(s_1)}(q_1)\z(q_2)\z(q_3)\rr}{\ll\g^{(+)}(q_1)\g^{(+)}(-q_1)\rr}\Big]_{HM} &=
\frac{24\sqrt{2}}{N^2\mathcal{N}_{(B)} q^3}\sin^2\vphi + O(q_1^2), \\[2ex]
%%%%%%%%%%%%%%%%%%%%%
\Big[\frac{\ll\g^{(s_1)}(q_1)\g^{(s_2)}(q_2)\z(q_3)\rr}{\ll\g^{(+)}(q_1)\g^{(+)}(-q_1)\rr}\Big]_{HM} &= 
-\frac{16}{N^2\mathcal{N}_{(A)}}\frac{q_1}{q^4}\cos\vphi\, (s_1s_2-\cos\vphi)^2+O(q_1^2),\\[2ex]
%%%%%%%%%%%%%%%%%%%%
\Big[\frac{\ll\g^{(s_1)}(q_1)\g^{(s_2)}(q_2)\g^{(s_3)}(q_3)\rr}{\ll\g^{(+)}(q_1)\g^{(+)}(-q_1)\rr}\Big]_{HM} &=
\frac{192\sqrt{2}}{N^2\mathcal{N}_{(A)}}\frac{1}{q^3}\sin^2\vphi \Big(\delta^{s_2s_3} 
+s_1(s_2+s_3)\frac{q_1}{q}\Big)+O(q_1^2).
\end{align}
}
Once again, we see these results are exactly in accordance with our consistency relations \eqref{cosmosoft1}-\eqref{cosmosoft6}, which therefore do indeed hold for the holographic model.

In fact, we are now in a position to understand a puzzling feature of the holographic model 3-point functions noted in section 7 of \cite{Bzowski:2011ab}. 
After defining dimensionless shape functions for general cosmological 3-point correlators, these shape functions were observed to have similar 
behaviour in the squeezed limit for both slow-roll inflation and the holographic model, with one curious exception: for slow-roll inflation the shape function $\mathcal{S}(\z\g^{(+)}\g^{(+)})$ has a simple pole as the momentum $q_1$ associated with the $\z$ vanishes, whereas its holographic model counterpart instead has a zero.

This discrepancy can now easily be understood using \eqref{cosmosoft3}.  Plugging this relation into the definitions of the shape function from \cite{Bzowski:2011ab}, in the squeezed limit $q_1\tto 0$ we find
\[\label{shape}
\mathcal{S}(\z\g^{(+)}\g^{(+)}) = -n_T(q)\frac{q_1^2}{q^2}\frac{\ll\z(q_1)\z(-q_1)\rr}{\ll\g^{(+)}(q)\g^{(+)}(-q)\rr}\big(1+O(q_1^2)\big).
\]
For slow-roll inflation we then have $\mathcal{S}_{SR}(\z\g^{(+)}\g^{(+)})=q/(2q_1)+O(q_1)$, but for the holographic model $\mathcal{S}_{HM}(\z\g^{(+)}\g^{(+)}) = q_1/(2q)+O(q_1^2)$ because the tensor tilt vanishes. 
As we have already emphasized, however, this vanishing tensor tilt for the holographic model is only an artifact of working to 1-loop order in the dual QFT: at 2-loops interactions typically generate a non-zero tensor tilt $n_T(q)\sim\geff^2$.
At 2-loop order then, the squeezed limit of the holographic model shape function will in fact be a pole after all, $\mathcal{S}_{HM}(\z\g^{(+)}\g^{(+)}) \sim \gYM^2N/q_1$.  
For the remaining polarisation $\z\g^{(+)}\g^{(-)}$ this issue does not arise, since the right-hand side of \eqref{cosmosoft3} vanishes identically.  Similarly, for the other correlators involving either one or three gravitons, the issue does not arise because for these correlators the 
corrections generated by small deviations from scale invariance are subleading, as we see from \eqref{cosmosoft2}-\eqref{cosmosoft6}.  The shape function for $\z\z\z$ was not examined in \cite{Bzowski:2011ab}, but in this case higher-loop corrections to the scalar tilt of the holographic model are clearly important.

\section{Discussion}\label{discussion}

The origin of the inflationary consistency relations in holographic cosmology is now clear.
By taking moments of the diffeomorphism Ward identity, we showed that correlators of the stress tensor in the dual QFT obey a set of non-perturbative soft theorems governing their behaviour in the limit as one momentum vanishes.  The soft theorems derived from the first and second moments are alone sufficient to fully determine the
leading and subleading soft behaviour.   Plugging this information into the holographic formulae connecting correlators of the dual QFT to bulk cosmological correlators, we immediately obtain the correct cosmological consistency relations to $O(q_1^2)$ for all 3-point functions of curvature perturbations and gravitons.
Besides furnishing a simple holographic derivation of the consistency relations, our analysis extends their validity to cosmologies in which the gravitational description is strongly coupled and only the dual QFT is tractable. 

The approach we have developed is both straightforward and systematic, and is easily generalised in a number of directions.  Firstly, as we will show elsewhere \cite{toappear}, the infinite set of higher-order consistency relations 
found in \cite{Hinterbichler:2013dpa} can be derived from the soft theorems associated with the third and higher moments of the diffeomorphism Ward identity in the dual QFT.  At these higher orders, it is no longer possible to fully undo the symmetrisation over indices that appears in the soft theorems, meaning that only partial constraints can be extracted instead of the complete higher-order soft behaviour.  Exactly the same was found on the cosmological side in \cite{Hinterbichler:2013dpa}.

Another obvious extension is to higher-point correlation functions, starting with the 4-point function (see, e.g., \cite{Seery:2006vu, Seery:2008ax, Kehagias:2012pd, Garriga:2013rpa, Ghosh:2014kba, Kundu:2014gxa}).  While the derivation of the necessary soft theorems for stress tensor correlators is straightforward, obtaining the holographic formulae connecting QFT and cosmological correlators is more involved.   Whatever method is used, we stress the importance of correctly determining the semilocal contact terms appearing in the numerators of the holographic formulae, without which it is impossible to recover the correct soft behaviour.  
(Indeed, this is ultimately the reason why it is necessary to keep track of such terms in the first place.) The importance of these terms was also emphasized in \cite{McFadden:2010vh, McFadden:2010jw, McFadden:2011kk, Bzowski:2011ab, Bzowski:2012ih}, where their contribution to local-type non-Gaussianity was noted. 
With an extension to higher-point functions in place, it would be interesting to examine internal \cite{Senatore:2012wy, Assassi:2012zq} and multiple soft limits \cite{Mirbabayi:2014zpa, Joyce:2014aqa} from a holographic perspective, as well as infrared loop effects \cite{Giddings:2010nc, Giddings:2010ui, Gerstenlauer:2011ti, Senatore:2012wy, Assassi:2012et, Senatore:2012nq, Pimentel:2012tw, Tanaka:2013caa}.

Finally, an important issue we have only touched on is the following. 
We know that the cosmological consistency relations can fail when any of their basic input assumptions are violated, e.g., by the presence of multiple scalar fields or non-Bunch-Davies initial conditions.  
Under such conditions, the holographic derivation of the consistency relations we have given must necessarily break down.  Understanding precisely how this occurs is an interesting direction for future work.  Nevertheless, our rough expectations 
are as follows.  Deviations from Bunch-Davies initial conditions for perturbations corresponds to evaluating dual QFT correlators in excited states instead of the Euclidean vacuum  \cite{McFadden:2009fg}.  In consequence, 1-point functions might not vanish in the absence of sources as assumed in our derivation of the Ward identities for stress tensor correlators.

In the case of cosmologies with multiple scalar fields, non-adiabatic backgrounds will generate spatially non-uniform sources for some of the scalar operators in the dual QFT.  This in turn produces extra contributions to the diffeomorphism Ward identity invalidating 
the soft theorems we have derived here. (Similar considerations 
arise in the wavefunction of the universe approach, as noted in \cite{Pimentel:2013gza}.)
For example, in the case of a single entropy mode, instead of \eqref{genrel} we would have
\[\label{new_Ward}
0=\nabla^i\<T_{ij}(\x)\>_s + \<\O(\x)\>_s\nabla_j\sigma(\x),
\]
where $\sigma(\x)$ is the spatially varying source for the entropy mode.  (We assume we can choose coordinates on field space so as to set the inflaton source to be spatially homogeneous in the background.)
Differentiating \eqref{new_Ward} with respect to the metric $(n{-}1)$ times before restoring the sources to their background values, we  obtain Ward identities mixing the stress tensor $n$-point function with the $n$-point function of one scalar and $(n{-}1)$-stress tensors (plus new contact terms).  Consequently, we can no longer simply relate the $n$-point function of the stress tensor to purely lower-point functions as we could before.
It would be instructive to study this in a specific setting, for example, the deformation of a CFT by multiple slightly relevant scalars in conformal perturbation theory \cite{Bzowski:2012ih, McFadden:2013ria, Garriga:2014fda}.

\bigskip

{\small  {\it Acknowledgments:}  Research at the Perimeter Institute is supported by the Government of Canada through Industry Canada and by the Province of Ontario through the Ministry of Research \& Innovation.  We thank also the UK Science \& Technology Facilities Council for support.}

\appendix

\section{Convergence of boundary terms}
\label{higher_app}

When we integrate by parts the $n$-th moment of the Ward identity \eqref{WardId}, we require the convergence of the boundary term\footnote{We can ignore the semilocal terms involving the $\Y_{ijkl}$ tensor as they only contribute when $\x_1$ coincides with $\x_2$ or $\x_3$, and so do not appear in the boundary term.}
\begin{align}\label{bdyterm}
\int \d^3\x_1\, \frac{\p}{\p x_{1i}}\Big[x_{1a_1}\ldots x_{1a_n} \<T_{ij}(\x_1)T_{kl}(\x_2)T_{mn}(\x_3)\>\Big].
\end{align}
In general, we expect this boundary term to be most singular for massless theories in which correlators 
decay algebraically rather than exponentially.
It is useful to consider the case of a conformal field theory.
As we take $\x_1$ to the boundary for finite $\x_2$ and $\x_3$ in the interior, we can use the operator product expansion to replace $T_{kl}(\x_2)T_{mn}(\x_3)$ with a single stress tensor insertion at $\x_2$ say.\footnote{Contributions to the OPE from other operators with different dimensions will vanish when inserted into the 2-point function with $T_{ij}(\x_1)$, while descendants of the stress tensor will give rise to less divergent behaviour as $x_1\tto\infty$.}
The resulting 2-point function $\< T_{ij}(\x_1)T_{ab}(\x_2)\>$ then scales as $x_1^{-6}$ as $x_1\tto \infty$, since the stress tensor has dimension three in three dimensions.  As the area of the boundary scales as $x_1^2$, we expect the boundary term \eqref{bdyterm} to converge for the first three moments, but a logarithmic divergence could potentially arise for the fourth.  
The existence of soft theorems for the fourth and higher moments 
is therefore non-trivial and the convergence of boundary terms should be checked on a case-by-case basis.\footnote{While the divergence of these higher moments can be regulated by the introduction of a Fourier transform factor $e^{-i\q_1\cdot\x_1}$ prior to the integration by parts, to deal with the extra terms thereby introduced requires additional smoothness assumptions as discussed in appendix \ref{momsp}.}  

Similar scaling arguments also apply to the evaluation of moments of  higher-point functions of the stress tensor, as would arise in the analysis of soft limits for higher-point cosmological correlators.  Applying the operator product expansion sequentially, we again expect the first three moments to converge, but the fourth and higher moments potentially diverge for massless QFTs. 
Such divergences, if present, would limit 
the amount of information we can extract about the soft limit.

\section{Momentum space}\label{Fourier_app}

%In this appendix we collect together a few auxiliary results relating to the stress tensor soft theorems in momentum space.

\subsection{Fourier transforms}\label{Fourier_details}

In this section we outline 
a few supplementary details concerning
the conversion of the position-space soft theorems \eqref{low1} and \eqref{low2} to their momentum-space counterparts \eqref{soft1} and \eqref{subsoft}.
The first step is to write the left-hand sides of \eqref{low1} and \eqref{low2} as
\begin{align}\label{introducingq}
\int\d^3\x_1\,\<T_{ij}(\x_1)\ldots\> &= \lim_{q_1\tto0}\int\d^3\x_1\,e^{-i\q_1\cdot\x_1}\<T_{ij}(\x_1)\ldots\>, \\
\int \d^3 \x_1\, x_{1(a}\<T_{b)j}(\x_1)\ldots\>
&= \lim_{q_1\tto 0} i\frac{\p}{\p q_{1(a}}\int \d^3\x_1\, e^{-i\q_1\cdot\x_1}\<T_{b)j}(\x_1)\ldots\>.
\end{align}
We then set $\x_3=-\x_2$, multiply by $8e^{-2i\q\cdot\x_2}$ and integrate over $\x_2$. 
Thus, for example,
\begin{align}
&\,\quad 8\int\d^3 \x_1\d^3 \x_2\, e^{-i\q_1\cdot\x_1 -2i\q\cdot\x_2} \,\<T_{ij}(\x_1)T_{kl}(\x_2)T_{mn}(-\x_2)\> \nn\\[1ex]
& = 8\int\d^3 \x_1\d^3 \x_2\d^3 \x_3 \, e^{-i\q_1\cdot\x_1 -i\q\cdot(\x_2-\x_3)}\,\<T_{ij}(\x_1)T_{kl}(\x_2)T_{mn}(\x_3)\>\delta(\x_2+\x_3)\nn\\[1ex]
&=8\int \frac{\d^3 \vec{p}}{(2\pi)^3}\int \d^3 \x_1\d^3 \x_2\d^3 \x_3 \, e^{-i\q_1\cdot\x_1 -i(\vec{p}+\q)\cdot\x_2-i(\vec{p}-\q)\cdot\x_3}\,\<T_{ij}(\x_1)T_{kl}(\x_2)T_{mn}(\x_3)\>\nn\\[1ex]
&= 8\int \frac{\d^3 \vec{p}}{(2\pi)^3}\, \<T_{ij}(\q_1)T_{kl}(\vec{p}+\q)T_{mn}(\vec{p}-\q)\>\nn\\[1ex]
&= 8\int \frac{\d^3 \vec{p}}{(2\pi)^3}\, \ll T_{ij}(\q_1)T_{kl}(\vec{p}+\q)T_{mn}(\vec{p}-\q)\rr (2\pi)^3\delta(\q_1+2\vec{p})\nn\\[1ex]
&=  \ll T_{ij}(\q_1)T_{kl}(\q_2)T_{mn}(\q_3)\rr,
\end{align}
with $\q_2$ and $\q_3$ fixed as given in \eqref{q2q3}.  Note the factor of $8$ cancels in the last line with the Jacobian factor from aligning the integration measure with the argument of the delta function.

The right-hand sides of \eqref{low1} and \eqref{low2} can be handled similarly by setting $\x_3=-\x_2$, multiplying by $8 e^{-2i\q\cdot\x_2}$ and integrating over $\x_2$.  For the last term on the right-hand side of \eqref{low1}, note we must evaluate the derivatives first before setting $\x_3=-\x_2$.  This can be accomplished, for example, by writing
\begin{align}
& \Big(x_{2i}\frac{\p}{\p x_{2j}}{+}x_{3i}\frac{\p}{\p x_{3j}}\Big)\<T_{kl}(\x_2)T_{mn}(\x_3)\> 
=-\int\frac{\d^3\vec{p}}{(2\pi)^3}\, e^{i\vec{p}\cdot(\x_2-\x_3)}\frac{\p}{\p p_i}\Big(p_j\ll T_{kl}(\vec{p})T_{mn}(-\vec{p})\rr\Big). 
\end{align}
For the last term on the right-hand side of \eqref{low2}, we have instead
\begin{align}
&\Big(x_{2a}x_{2b}\frac{\p}{\p x_{2j}}{+}x_{3a}x_{3b}\frac{\p}{\p x_{3j}}\Big)\<T_{kl}(\x_2)T_{mn}(\x_3)\> \nn\\& = (x_{2a}x_{2b}-x_{3a}x_{3b})\int\frac{\d^3 \vec{p}}{(2\pi)^3}\,e^{i\vec{p}\cdot(\x_2-\x_3)}\ll T_{kl}(\vec{p})T_{mn}(-\vec{p})\rr,
\end{align}
which vanishes when we set $\x_3=-\x_2$. 
For this reason the right-hand side of \eqref{subsoft} contains only single derivatives with respect to momenta, rather than double as might have been expected.

\subsection{Differentiating the momentum-space Ward identity}
\label{momsp}

An alternative route to obtain the soft theorems \eqref{soft1} and \eqref{soft2} is to start directly from the 3-point Ward identity \eqref{WardId} in momentum space, which reads
\begin{align}\label{mom_space_WardId}
& q_{1i}\Big[\<\!\<T_{ij}(\q_1)T_{kl}(\q_2)T_{mn}(\q_3)\>\!\>\nn\\&\qquad
-2\ll T_{ij}(\q_1)\Y_{klmn}(\q_2,\q_3)\rr
-2\<\!\<\Upsilon_{ijmn}(\q_1,\q_3)T_{kl}(\q_2)\>\!\>
-2\<\!\<\Upsilon_{ijkl}(\q_1,\q_2)T_{mn}(\q_3)\>\!\>\Big] \nn\\[1ex]&=
2q_{1(k}\<\!\<T_{l)j}(\q_3)T_{mn}(-\q_3)\>\!\>
+2q_{1(m}\<\!\<T_{n)j}(\q_2)T_{kl}(-\q_2)\>\!\>
+\delta_{kl}q_{2 p}\<\!\<T_{pj}(\q_3)T_{mn}(-\q_3)\>\!\>\nn\\[1ex]&\quad
+\delta_{mn}q_{3 p}\<\!\<T_{pj}(\q_2)T_{kl}(-\q_2)\>\!\>
-q_{2j}\<\!\<T_{kl}(\q_3)T_{mn}(-\q_3)\>\!\>
-q_{3j}\<\!\<T_{mn}(\q_2)T_{kl}(-\q_2)\>\!\>.
\end{align}
To proceed we need to assume the squeezed limit is smooth in the sense that both the Ward identity \eqref{mom_space_WardId} and the stress tensor 3-point function (along with the semilocal terms inside the square bracket above) are at least twice differentiable in the limit as $q_1\tto 0$.
Setting $\q_2$ and $\q_3$ as in \eqref{q2q3} and  differentiating \eqref{mom_space_WardId} with respect to $\q_{1}$ (using the chain rule where appropriate), in the limit $q_1\tto 0$ we do indeed recover precisely the leading soft theorem \eqref{soft1}.  Notice here that the combination  
\begin{align}\label{extra_term}
&\lim_{q_1\tto 0}q_{1i}\frac{\p}{\p q_{1a}}\Big[\ll T_{ij}(\q_1)T_{kl}(\q_2)T_{mn}(\q_3)\rr \nn\\[1ex]&\quad
-2\ll T_{ij}(\q_1)\Y_{klmn}(\q_2,\q_3)\rr
-2\<\!\<\Upsilon_{ijmn}(\q_1,\q_3)T_{kl}(\q_2)\>\!\>
-2\<\!\<\Upsilon_{ijkl}(\q_1,\q_2)T_{mn}(\q_3)\>\!\>\Big]
\end{align}
vanishes since differentiability implies that the derivative of the stress tensor 3-point function (plus semilocal terms) is finite in the limit $q_1\tto 0$, i.e., there are no poles to counteract the overall factor of $q_{1i}$.

Similarly, to recover the subleading soft theorem \eqref{subsoft}, we impose \eqref{q2q3} and differentiate \eqref{mom_space_WardId} twice with respect to $\q_1$ before taking the limit $q_1\tto 0$.
This procedure yields exactly \eqref{subsoft},  noting that the combination 
\begin{align}\label{extra2}
&\lim_{q_1\tto 0} q_{1i}\frac{\p}{\p q_{1a}}\frac{\p}{\p q_{1b}}\Big[\ll T_{ij}(\q_1)T_{kl}(\q_2)T_{mn}(\q_3)\rr
\nn\\[1ex]&\quad
-2\ll T_{ij}(\q_1)\Y_{klmn}(\q_2,\q_3)\rr
 -2\<\!\<\Upsilon_{ijmn}(\q_1,\q_3)T_{kl}(\q_2)\>\!\>
-2\<\!\<\Upsilon_{ijkl}(\q_1,\q_2)T_{mn}(\q_3)\>\!\>\Big]
\end{align}
vanishes, again by the assumed differentiability of the stress tensor 3-point function plus semilocal terms in the limit $q_1\tto 0$.

At first sight, it seems puzzling that we have been able to recover the complete stress tensor 3-point function up to terms of order $q_1^2$ when solutions of the Ward identity \eqref{mom_space_WardId} are ambiguous up to the addition of a transverse term. 
Such terms are tacitly forbidden, however, by our assumption that 
the stress tensor 3-point function (plus appropriate semilocal terms) is twice differentiable in the limit as $q_1\tto 0$.  
To see this, notice that the general form of any ambiguous contribution\footnote{See appendix A.1 of \cite{Bzowski:2013sza} for a complete classification.}
 is $\pi_{a(i}(\q_1)\pi_{j)b}(\q_1)X_{abklmn}(\q_1,\q)$.  
The action of two derivatives with respect to $\q_1$ on the projection operators then produces a double pole, as can be seen from \eqref{pi_deriv}.  For this to cancel requires $X_{abklmn} = O(q_1^2)$, rendering the stress tensor 3-point function unambiguous to the required order.

The smoothness assumption we have employed in this momentum-space derivation appears to be unnecessary in the position-space approach we used in the main text.   The origin of this subtle difference can be seen as follows.  In our position-space approach, we took the route
\[
\int\d^3 \x_1\, x_{1a}\frac{\p}{\p x_{1i}}\<T_{ij}(\x_1)\ldots\> = -\int \d^3 \x_1\,\<T_{aj}(\x_1)\ldots\>
= -\lim_{q_1\tto 0}\int\d^3\x_1\, e^{-i\q_1\cdot\x_1}\<T_{aj}(\x_1)\ldots\>
\]
whereas the momentum-space approach above is equivalent to
\[
\lim_{q_1\tto 0} \frac{\p}{\p q_{1a}}\Big[ q_{1i}\<T_{ij}(\q_1)\ldots\>\Big]
=\lim_{q_1\tto 0}\int\d^3\x_1\, e^{-i\q_1\cdot\x_1}x_{1a}\frac{\p}{\p x_{1i}}\<T_{ij}(\x_1)\ldots\>.
\]
Thus, in the position-space approach, the integration by parts is performed first, before the exponential factor is introduced.  In the momentum-space approach, however, the exponential factor is introduced prior to the integration by parts, generating the extra terms \eqref{extra_term} and \eqref{extra2} above.  Our smoothness assumption was then required to guarantee the vanishing of these terms.

\vspace{-2mm}

\section{Helicity basis}
\label{helicity_basis_app}

This appendix provides further details about the helicity basis we adopt in the main text. In section \ref{helicity_app}, we define our conventions and list the assorted contractions of polarisation tensors that feature in the holographic formulae \eqref{holo_zzz}-\eqref{holo_ggg}. In section \ref{momderiv}, we discuss the differentiation of polarisation tensors with respect to momentum.

\vspace{-1mm}
\subsection{Conventions}\label{helicity_app}

Our polarisation tensors $\ep_{ij}^{(s)}(\q)$ satisfy
\[
\ep_{ij}^{(s)}(\q)=\ep_{ji}^{(s)}(\q), \qquad \ep_{ii}^{(s)}(\q)=0,\qquad q_i\ep_{ij}^{(s)}(\q)=0,
\]
and are normalised such that \cite{WeinbergBook}
\[
\label{PiTTdecomp}
\Pi_{ijkl}(\q) = \half\ep^{(s)}_{ij}(\q)\ep^{(s)}_{kl}(-\q), \qquad
 \ep^{(s)}_{ij}(\q)\ep^{(s')}_{ij}(-\q) = 2\delta^{ss'},
\]
where the transverse traceless projector $\Pi_{ijkl}$ was defined in \eqref{projection_operators}.
Helicity indices $s$, $s'$, etc., take values $\pm 1$ and we sum over repeated indices.  

When dealing with 3-point functions, momentum conservation implies that all three momenta lie in a single plane.  Taking this plane to be the $(x,z)$ plane, for some momentum $\q = q\,(\sin\theta,\,0,\,\cos\theta)$ we then have
\begin{align}
\label{helicity_basis}
 \ep_{ij}^{(s)}(\q) = 
 \frac{1}{\sqrt{2}}\left(\begin{array}{ccc} \cos^2\theta\,\,\,\, & is\cos\theta \,\,\,\,& -\sin\theta\cos\theta \\ is\cos\theta\,\,\,\, & -1\,\,\,\, & -is\sin\theta \\ -\sin\theta\cos\theta\,\,\,\, & -is\sin\theta \,\,\,\,& \sin^2\theta \end{array}\right). 
\end{align}
Using this representation we can evaluate contractions of polarisation tensors such as those that appear in the holographic formulae
\eqref{holo_zzz}-\eqref{holo_ggg}, namely
\begin{align}
\label{Theta_def}
&\Theta_1^{(s_3)} = \pi_{ij}(\q_1)\ep_{ij}^{(s_3)}(-\q_3), \qquad
&& \Theta_2^{(s_3)} = \pi_{ij}(\q_2)\ep_{ij}^{(s_3)}(-\q_3), \nn\\
& \Theta^{(s_2s_3)} = \pi_{ij}(\q_1)\ep_{ik}^{(s_2)}(-\q_2)\ep_{kj}^{(s_3)}(-\q_3), \qquad
&&  \theta^{(s_2s_3)} = \ep^{(s_2)}_{ij}(-\q_2)\ep^{(s_3)}_{ij}(-\q_3), \nn\\
&\Theta^{(s_1s_2s_3)} = \ep^{(s_1)}_{ij}(-\q_1)\ep^{(s_2)}_{jk}(-\q_2)\ep^{(s_3)}_{ki}(-\q_3),
\end{align}
where the transverse projection operator $\pi_{ij}$ was defined in \eqref{PiTTdecomp}. 
In terms of the symmetric polynomials of momentum magnitudes defined in \eqref{abc_def} and \eqref{lambda_def}, we find\footnote{See appendix C of \cite{McFadden:2011kk} and appendix A of \cite{Bzowski:2011ab} for further details.}
\begin{align}
\label{Theta_results}
&\Theta_1^{(\pm)} = -\frac{\lambda^2}{4\sqrt{2} b_{13}^2},
&&\hspace{-4.5cm}\Theta_2^{(\pm)} = -\frac{\lambda^2}{4\sqrt{2} b_{23}^2},\nn\\[1ex]
&\Theta^{(+++)} = -\frac{\lambda^2 a_{123}^2}{16\sqrt{2}c_{123}^2},
&&\hspace{-4.5cm}\Theta^{(++-)} = -\frac{\lambda^2}{16\sqrt{2}c_{123}^2}(q_3-a_{12})^2, \nn\\[1ex]
&\theta^{(++)} = \frac{a_{123}^2(a_{23}-q_1)^2}{8b_{23}^2},
&&\hspace{-4.5cm}\theta^{(+-)} = \frac{(a_{13}-q_2)^2(a_{12}-q_3)^2}{8b_{23}^2}, \nn\\[1ex]
&\Theta^{(++)}  =
\frac{a_{123}(a_{23}-q_1)}{16c_{123}^2}\,\big[2q_1^2 a_{123}(a_{23}-q_1)-\lambda^2\big],&&  \nn\\[1ex]
&\Theta^{(+-)} =
 \frac{(a_{13}-q_2)(a_{12}-q_3)}{16c_{123}^2}\,\big[2q_1^2(a_{13}-q_2)(a_{12}-q_3)+\lambda^2\big]. &&
\end{align}
The representation \eqref{helicity_basis} can also be used to evaluate the contractions of polarisation tensors arising when converting our soft theorems to a helicity basis in section \ref{hel_conv}.

\subsection{Derivative of a polarisation tensor with respect to momentum}\label{momderiv}

We now turn to the derivation of \eqref{ep_deriv}, namely
\[\label{ep_deriv2}
\frac{\p}{\p q_a}\ep_{ij}^{(s)}(\q)=-\frac{2}{q^2}q_{(i}\ep_{j)a}^{(s)}(\q).
\]
One method is to note that the most general form the right-hand side could take is
\[
\frac{\p}{\p q_a}\ep_{ij}^{(s)}(\q)= A_1(q) q_a \delta_{ij}+A_2(q) q_{(i}\delta_{j)a}+A_3(q)q_aq_iq_j+A_4(q)q_a\ep^{(s)}_{ij}(\q)+A_5(q)q_{(i}\ep_{j)a}^{(s)}(\q),
\]
for some unknown scalar coefficients $A_n(q)$.  We then enforce, e.g., 
\begin{align}
q_i\frac{\p}{\p q_a}\ep_{ij}^{(s)}(\q)=-\ep_{ja}^{(s)}(\q),\qquad
\delta_{ij}\frac{\p}{\p q_a}\ep_{ij}^{(s)}(\q)=0, \qquad
q_a\frac{\p}{\p q_a}\ep_{ij}^{(s)}(\q)=0, 
\end{align}
where the first two relations encode the transverse tracelessness of $\ep_{ij}^{(s)}(q)$ and the last arises from its independence under rescalings of $\q$.  Satisfying these relations requires $A_1(q)=A_2(q)=A_3(q)=A_4(q)=0$ and $A_5(q)=-2/q^2$, yielding \eqref{ep_deriv2}.

\begin{figure}[t]
\center
\includegraphics[width=1.8cm]{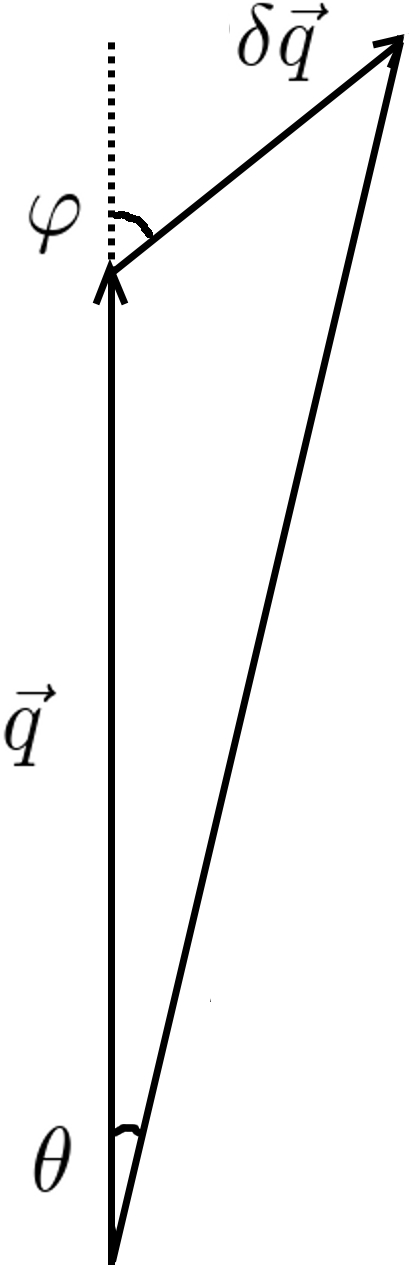}
\caption{\label{ep_fig}
Given arbitrary vectors $\q$ and $\delta\q$ we can orient our coordinates such that \eqref{orientation} holds.}
\end{figure}

As a check on this relation, 
consider two arbitrary vectors $\q$ and $\delta \q$ where, without loss of generality, we can choose a Cartesian coordinate system such that
\[\label{orientation}
\q = q(0,\,0,\,1), \qquad \delta\q = \delta q (\sin\vphi,\,0,\,\cos\vphi),\qquad
\q+\delta\q = (q+\delta q)(\sin\theta,\,0,\,\cos\theta),
\]
as illustrated in Figure \ref{ep_fig}.  Here $\theta = (\delta q/q) \sin\vphi+O(\delta q^2)$ is small, but $\vphi$ is not necessarily so. 
We then have
\begin{align}\label{exp1}
 \ep^{(s)}(\q+\delta \q) &= \frac{1}{\sqrt{2}}\left(\begin{array}{ccc} \cos^2\theta\,\,\,\, & is\cos\theta \,\,\,\,& -\sin\theta\cos\theta \\ is\cos\theta\,\,\,\, & -1\,\,\,\, & -is\sin\theta \\ -\sin\theta\cos\theta\,\,\,\, & -is\sin\theta \,\,\,\,& \sin^2\theta \end{array}\right) \nn\\&
 = \ep^{(s)}(\q)+\frac{\delta q}{\sqrt{2}q}\sin\vphi \left(\begin{array}{ccc} 0 & 0 & -1 \\ 0&0&-is\\-1&-is&0
  \end{array}\right)+O(\delta q^2).
 \end{align}
For \eqref{ep_deriv2} to be valid, this must match
\[
\ep_{ij}^{(s)}(\q+\delta\q)=\ep_{ij}^{(s)}(\q)+\delta q_a\frac{\p}{\p q_a}\ep_{ij}^{(s)}(\q)+O(\delta q^2)
=\ep_{ij}^{(s)}(\q)-\frac{2}{q^2} q_{(i}\ep_{j)a}^{(s)}(\q)\delta q_a+O(\delta q^2).
\]
Since 
\begin{align}
\ep^{(s)}(\q) = \frac{1}{\sqrt{2}}\left(\begin{array}{ccc} 1&is&0\\ is&-1&0\\ 0&0&0\end{array}\right),
\end{align}
we have
\[
-\frac{2}{q^2}q_{(i}\ep_{j)a}^{(s)}(\q)\delta q_a = 
-\frac{2}{q}\, \delta_{z(i}\ep_{j)x}^{(s)}(\q)\delta q \sin\vphi,
\]
which indeed reproduces \eqref{exp1}.

%\vspace{-2mm}

\bibliography{Soft}
\bibliographystyle{jhep}

\end{document}